\documentclass[twocolumn]{aastex631}
\usepackage{amsmath}
\usepackage{mathtools}

\newcommand\gaia{\textit{Gaia}}

\newcommand{\llnl}{Space Science Institute, Lawrence Livermore National Laboratory, 7000 East Ave., Livermore, CA 94550, USA}
\newcommand{\UCB}{University of California, Berkeley, Astronomy Department, Berkeley, CA 94720, USA}
\newcommand{\UCR}{Department of Physics and Astronomy, University of California, Riverside, 900 University Avenue, Riverside, CA 92521, USA}
\newcommand{\UM}{Department of Physics, University of Michigan, 450 Church St, Ann Arbor, MI 48109, USA}
\newcommand{\LCTP}{Leinweber Center for Theoretical Physics, 450 Church St, Ann Arbor, MI 48109, USA}

\begin{document}

\title{On Finding Black Holes in Photometric Microlensing Surveys}

\author[0009-0007-4089-5012]{Zofia Kaczmarek\,*}
\email{*zofia.kaczmarek@uni-heidelberg.de}
\affiliation{\llnl}
\affiliation{Zentrum f{\"u}r Astronomie der Universit{\"a}t Heidelberg, Astronomisches Rechen-Institut, M{\"o}nchhofstr. 12-14, 69120 Heidelberg, Germany}
\author[0000-0002-1052-6749]{Peter McGill\,*}
\email{*mcgill5@llnl.gov}
\affiliation{\llnl}
\author[0000-0002-5910-3114]{Scott E. Perkins\,}
\affiliation{\llnl}
\author[0000-0003-0248-6123]{William A. Dawson\,}
\affiliation{\llnl}
\author[0000-0003-4591-3201]{Macy Huston}
\affiliation{\UCB}
\author[0000-0002-4457-890X]{Ming-Feng Ho\,}
\affiliation{\UCR}
\affiliation{\UM}
\affiliation{\LCTP}
\author[0000-0002-0287-3783]{Natasha S. Abrams}
\affiliation{\UCB}
\author[0000-0001-9611-0009]{Jessica R. Lu}
\affiliation{\UCB}

\begin{abstract}

There are expected to be millions of isolated black holes in the Galaxy resulting from the death of massive stars. Measuring the abundance and properties of this remnant population would shed light on the end stages of stellar evolution and the evolution paths of black hole systems. Detecting isolated black holes is currently only possible via gravitational microlensing which has so far yielded one definitive detection. The difficulty in finding microlensing black holes lies in having to choose a small subset of events based on characteristics of their lightcurves to allocate expensive and scarce follow-up resources to confirm the identity of the lens. Current methods either rely on simple cuts in parameter space without using the full distribution information or are only effective on a small subsets of events. In this paper we present a new lens classification method. The classifier takes in posterior constraints on lightcurve parameters and combines them with a Galactic simulation to estimate the lens class probability. This method is flexible and can be used with any set of microlensing lightcurve parameters making it applicable to large samples of events. We make this classification framework available via the {\tt popclass} python package. We apply the classifier to $\sim10,000$ microlensing events from the OGLE survey and find $23$ high-probability black hole candidates. Our classifier also suggests that the only known isolated black hole is an observational outlier according to current Galactic models and allocation of astrometric follow-up on this event was a high-risk strategy. 

\end{abstract}


\section{Introduction} \label{sec:intro}

Gravitational microlensing is sensitive to both luminous and dark objects, making it a powerful tool for studying a variety of Galactic populations (from exoplanets; e.g., \citealt{Ryu2018}, to white dwarfs; \citealt{Sahu2017, McGill2018,McGill2023}, to black holes; e.g., \citealt{Lu2016,Sahu2022, Lam2022, Lam2023} and stars e.g., \citealt{Kluter2020, McGill2020}). Microlensing's broad power has been realized in the era of modern variability surveys such as The Optical Gravitational Lensing Experiment \citep[OGLE;][]{OGLE-IV}, Microlensing Observations in Astrophysics \citep[MOA;][]{Hearnshaw2006,Sumi2013} survey, The Korea Microlensing Telescope Network \citep[KMTNet;][]{KMTNet} and The VISTA Variables in The Via Lactea \citep[VVV;][]{Minniti2010} survey that have provided databases containing of order ten thousand microlensing events \citep[e.g.,][]{Mroz2020, Husseiniova2021, Shin2024}. This yield of events is only set to increase with the advent of the Roman Space Telescope \citep[][]{Penny2019, Johnson2020} unlocking high Galactic Bugle event rates in the near-infrared \citep[e.g.,][]{Gould1994, Shvartzvald2017, McGill2019, Kaczmarek2024} and the Vera C. Rubin Observatory \citep{Ivezic2019,Abrams2023}.

One of the promised strengths of microlensing is its unique sensitivity to dark and isolated objects, specifically the expected population of stellar-origin black holes (SOBHs) in the Galaxy \citep[e.g.,][]{Mao2002, Agolfinding2002}. As SOBHs are the terminal evolutionary state of massive stars, the Milky Way is expected to contain $10^8 - 10^9$ of them \citep{AgKa2002}. Despite this large expected abundance, only $\sim50$ SOBHs are known, the majority of which reside in X-ray binaries \citep[e.g.,][]{CygnusX11, CygnusX12, BHbinaries1, BHbinaries2} with others in astrometric or spectroscopic binaries \citep{GaiaBH1, GaiaBH2, GaiaBH3}. In addition to known Galactic population of SOBHs, $\sim 10^2$ extra-galactic black holes, spanning masses from 2-200$M_\odot$, have been observed in binary black hole or neutron star - black hole mergers causing gravitational waves \citep{GW1, GW2, GW3, GW4, GW5}. However, all current black hole detection methods, apart from microlensing, are only sensitive to black holes in multiple systems. 

Compiling a dataset of isolated SOBHs detected via microlensing would enable investigations of SOBHs beyond the current sample that followed an evolutionary path resulting in a stable binary system. A sample of isolated SOBHs could answer open questions such as if the Galactic and extragalactic mass spectrum of dark remnants are similar, or whether the mass gap observed between neutron stars and black holes results from observational biases \citep[e.g.,][]{WyrzMandel2020} or an astrophysical mechanism \citep[e.g.,][]{Farr2011, Kreidberg2012, Belczynski2012, Shao2022, GW5, Barr2024}. 

The first and only current detection of an isolated black hole was recently reported via microlensing event OGLE-2011-BLG-0462/MOA-2011-BLG-191 \citep{Sahu2022, Lam2022, Lam2023}. However, determining that the lens of this event was a black hole was not straightforward. This difficulty arises because the photometric microlensing signal is degenerate in lens-source physical parameters -- more information is needed to break these degeneracies \citep[e.g.,][]{Rybicki2018}. Astrometric follow-up from the Hubble Space Telescope was required to measure the event's astrometric microlensing signal, which was critical to identifying the lens as a black hole. In this common scenario, using characteristics of the photometric signal, the lens had to be identified as a likely black hole candidate to justify the use of expensive follow-up observations to confirm its nature. 

Even in the upcoming era of multiple space-based sub-milliarcsecond (mas) capable observatories (e.g., the James Webb Space Telescope; \citealt{Gardner2006}, the Roman Space Telescope; \citealt{Spergel2015}, and \gaia; \citealt{Prusti2016}), the majority of microlensing events will only be detected photometrically in the first instance. Therefore, determining if a given event is caused by a black hole based on only photometric signals remains critical to efficiently allocate expensive follow-up resources (e.g., further photometric or astrometric observations via adaptive optics; \citealt{Terry2022}, interferometry; \citealt{Dong2019}, or late-time high resolution imaging; \citealt{Abdurrahman2021}) to increase the yield of detected isolated black holes. 

Microlensing occurs when an intervening massive object gravitationally deflects and magnifies the light from a more distance background source. For the majority of microlensing events the only information available from the lightcurve that traces physical parameters of the lens and is almost always well constrained is the Einstein timescale of the event,

\begin{equation}
    t_{\rm E} = \frac{D_{L}\theta_{\rm E}}{v_{\rm rel}}.
    \label{eq:tE}
\end{equation}

Here, $v_{\rm rel}$ is the relative lens-source transverse velocity, $D_{L}$ is the distance to the lens, $D_{S}$ is the distance to the source, and $t_{\rm E}$ is the time it takes to cross the angular Einstein radius of the system $\theta_{\rm E}=\sqrt{4GM_{L}c^{-2}(D^{-1}_{L}-D^{-1}_{S})}$, where $M_{L}$ is the mass of the lens. 

For some events, the imprint of Earth's orbital acceleration can be detected as an asymmetrical imprint on the microlensing light curve \citep[e.g.,][]{Wyrzykowski2015, Kaczmarek2022} allowing microlensing parallax,

\begin{equation}
    \pi_{\rm E} = \frac{1 \ {\rm AU}}{\theta_{\rm E}}\left(\frac{1}{D_{L}}-\frac{1}{D_{S}}\right),
    \label{eq:piE}
\end{equation}
to be well-constrained. For events without a clear parallax signal, upper limits can be placed on $\pi_{\rm E}$ \citep[e.g.,][]{golovich}. However, these parameters alone are not sufficient to directly determine the lens mass and identify nature of the lens due to various degeneracies \citep[e.g.,][]{Sm03, Gould2004}. 

To overcome this challenge, Galactic and Stellar models can be used in combination with microlensing lightcurve constraints to infer the missing lens parameters. The state-of-the-art implementations of this approach are detailed in \citet{Howil2024} and \cite{Bachelet2024} and implemented in the {\tt DarkLensCode}\footnote{\url{https://github.com/BHTOM-Team/DarkLensCode}} and {\tt pyLIMass} python packages, respectively. {\tt DarkLensCode} has been applied to samples of events with large parallax signals and has found many promising black hole candidates \citep{Wyrzykowski2016, Kaczmarek2022,Kruszynska2024,Rybicki2024}. {\tt pyLIMass} can use all available data for a microlensing event and recovers lens masses with a median precision of $20\%$ over a simulated {\it Roman} dataset \citep{Bachelet2024}.

{\tt DarkLensCode} is designed to find black hole lenses and works by leveraging a non-zero constraint on $\pi_{\rm E}$ along with a source proper motion estimate to constrain the lens motion which is crucial input to a Galactic phase-space density model. {\tt DarkLensCode} uses this information in combination with the fraction of blended light in the event to derive lens mass and distance estimates along with the probability that the lens is dark, i.e., the probability that the fraction of blended light cannot be explained by a stellar lens at a consistent distance. While this approach proves effective for events with a nearby lens, large parallax signal, and auxiliary proper motion information, it is not designed to classify the majority of events that do not have these characteristics.

In this work, we develop a complimentary method to \cite{Howil2024} to find black hole lenses that fills the gap in microlensing dark remnant search and classifications. We build a Bayesian classifier that takes posterior samples from lightcurve event modelling as input and returns the probability of the event belonging to a given astrophysical class given a Galactic model. This classification method is flexible and can be used with any Monte Carlo simulation of the Galaxy permitting straightforward ways to test the effect of different Galactic model assumptions on lens classification. The classifier does not require auxiliary information such as source astrometry, and can be used in weak parallax regimes where most SOBH lenses are expected to reside \citep[e.g.,][]{Lam2020}. Overall this means the classifier can be quickly and uniformly applied to entire catalogs of photometric microlensing events.

The paper is structured as follows. In Section \ref{sec:data} we describe the data used in this work, comprising of posterior samples from \citet{golovich} of parameters of $\sim10,000$ OGLE-III and OGLE-IV microlensing events \citep{Wyrzykowski2015, Mroz2019}.
In Section \ref{sec:classification}, we detail our lens classification method. In particular, we describe the specific Galactic simulations used by the classifier and detail the classification procedure based on Bayesian statistics. In Section \ref{sec:classification_results} we present the classification results. In particular, we compare classification with different underlying initial-final mass relations, we present a sample of 23 high-probability black hole candidates and outline a fast method of approximating $p(\textnormal{SOBH} | \boldsymbol{d},\mathcal{G})$, the probability of the lens belonging to the stellar-origin black hole class. We also discuss the candidates in detail, including their expected astrometric signal. Finally, in Section \ref{sec:conclusions} we discuss and summarise implications of this work for upcoming surveys such as the {\it Vera C. Rubin Observatory} and the {\it Roman Space Telescope}. 

\section{Data}
\label{sec:data}

We use the dataset of publicly-available microlensing events from OGLE-III \citep{OGLEIII} and IV \citep{OGLE-IV} detected towards the Galactic bulge. The OGLE-III and IV surveys were conducted at the 1.3-meter Warsaw University Telescope, Las Campanas Observatory in Chile between 2002-2009 and 2010-2017, respectively.
Specifically, we use the set of 3560 microlensing events found by \citet{Wyrzykowski2015} and the set of 5790 low-cadence field events in \citet{Mroz2019}\footnote{We omit the high-cadence field data from \cite{Mroz2019} due to this data not being publicly available.}.
For this total set of 9350 events, we use the posterior distributions of microlensing event parameters obtained in \citet{golovich}\footnote{The posterior samples obtained from this analysis are available at \url{https://gdo-microlensing.llnl.gov}}. 
\citet{golovich} simultaneously modelled microlensing parallax, systematic instrumental effects and source variability, to obtain posterior distributions that minimized bias in the physical parameters of the events.

\section{Bayesian Lens Classification} \label{sec:classification}

\subsection{Galactic Model}

For the model of the Galaxy and simulation of microlensing events, $\mathcal{G}$, we use the Population Synthesis Code for Compact Object Microlensing Events \citep[\texttt{PopSyCLE};][]{Lam2020}. The underlying stellar population in \texttt{PopSyCLE} uses the the Besançon model \citep{Robin2004} implemented by \texttt{Galaxia} \citep{Sharma2011}. In this simulation, SOBHs, Neutron Stars (NSs), and White Dwarfs (WDs) are generated by evolving clusters matching thin and thick
disk, bulge, stellar halo stellar populations in \texttt{Galaxia} via the Population Interface for Stellar Evolution and Atmospheres code \citep[\texttt{SPISEA};][]{Hosek2020}. 

\begin{table}
\begin{center}
\begin{tabular}{p{40mm}p{38mm}}
    \hline\hline 
    Parameter & Value \\\hline
    Milky Way escape velocity & $550$kms$^{-1}$ \citep{Piffl2014} \\
    Sun-Galactic center distance & $8.3$kpc \\
    Peak of initial SOBH progenitor kick distribution & $100$kms$^{-1}$ \\
    Peak of initial NS progenitor kick distribution & $350$kms$^{-1}$ \\
    Initial-Final Mass Relation & \{\citet{Sukhbold2016}, \citet{Raithel2018}, \citet{Spera2015}\}\\
    Extinction Law & \citet{Damineli2016} \\
    Bar dimensions (radius, major axis, minor axis, height) & $(2.54, 0.70, 0.424, 0.424)$ kpc \\
    Bar angle (Sun–Galactic center, 2nd, 3rd) & $(62.0, 3.5, 91.3)$ $^{\circ}$ \\
    Bulge velocity dispersion (radial, azimuthal, z) & $(100, 100, 100)$ kms$^{-1}$ \\
    Bar patternspeed & 40.00kms$^{-1}$ kpc$^{-1}$ \\
    multiplicity & singles \\
    \hline
\end{tabular}
\end{center}
\caption{\label{tab:simulation} \texttt{PopSyCLE} simulation parameters. The implementation of the IFMR is described in detail in \cite{Rose2022}. Galactic parameters are consistent with ''v3'' in \citet[App.~A]{Lam2020}, and match the event rates reported by OGLE~\citep{Mroz2019}. We investigate three different options for the IFMRs.}
\end{table}

\texttt{PopSyCLE} breaks up the \texttt{Galaxia} population into age and metallicity bins and \texttt{SPISEA} generates single-age, single-metallicity and single-IMF clusters to match the binned distributions. \texttt{SPISEA} uses an initial mass function, stellar multiplicity, extinction law, metallicity-dependent stellar evolution, and a separate initial final mass relation for SOBHs, NSs and WDs \citep{Lam2020, Kalirai2008}. Finally, SOBHs and NSs are given initial progenitor kick velocities. All values and relationships adopted for our simulations are in Table \ref{tab:simulation}. For this study we explore Galactic models with three different IFMRs \citep{Sukhbold2016, Raithel2018, Spera2015} because this is likely to affect the predictions for SOBHs in our classification framework \citep{Rose2022}. Hereafter, we use the \citet{Sukhbold2016} IFMR as the default choice of model (e.g. for visualization), as it is the most updated with recent results and includes both metallicity dependence and explosion physics. Still, we report classification results with all IFMRs, and we consider all IFMRs equally for BH candidate selection.

To generate a representative sample of microlensing events, we ran the Galactic simulation over $20$ different locations across the OGLE bulge fields with an area of $0.3$ square degrees each (see Table \ref{tab:sim_stats} in Appendix \ref{app:sim}). Microlensing events are selected from this simulation by selecting lens-source pairs that come within $\theta_{\rm E}$ separation of each other and have a source I-band baseline magnitude $< 21$. We note that OGLE survey selection functions (e.g., $t_{\rm E}$ selection efficiency) do not need to be included when calculating the class of single microlensing events -- see section 3.2 of \cite{Perkins2024} for a derivation of this point. 

\subsection{Classification procedure}\label{sec:bayes}

\begin{figure}
    \centering
    \includegraphics[width=\columnwidth, trim=0cm 0cm 0cm 0cm]{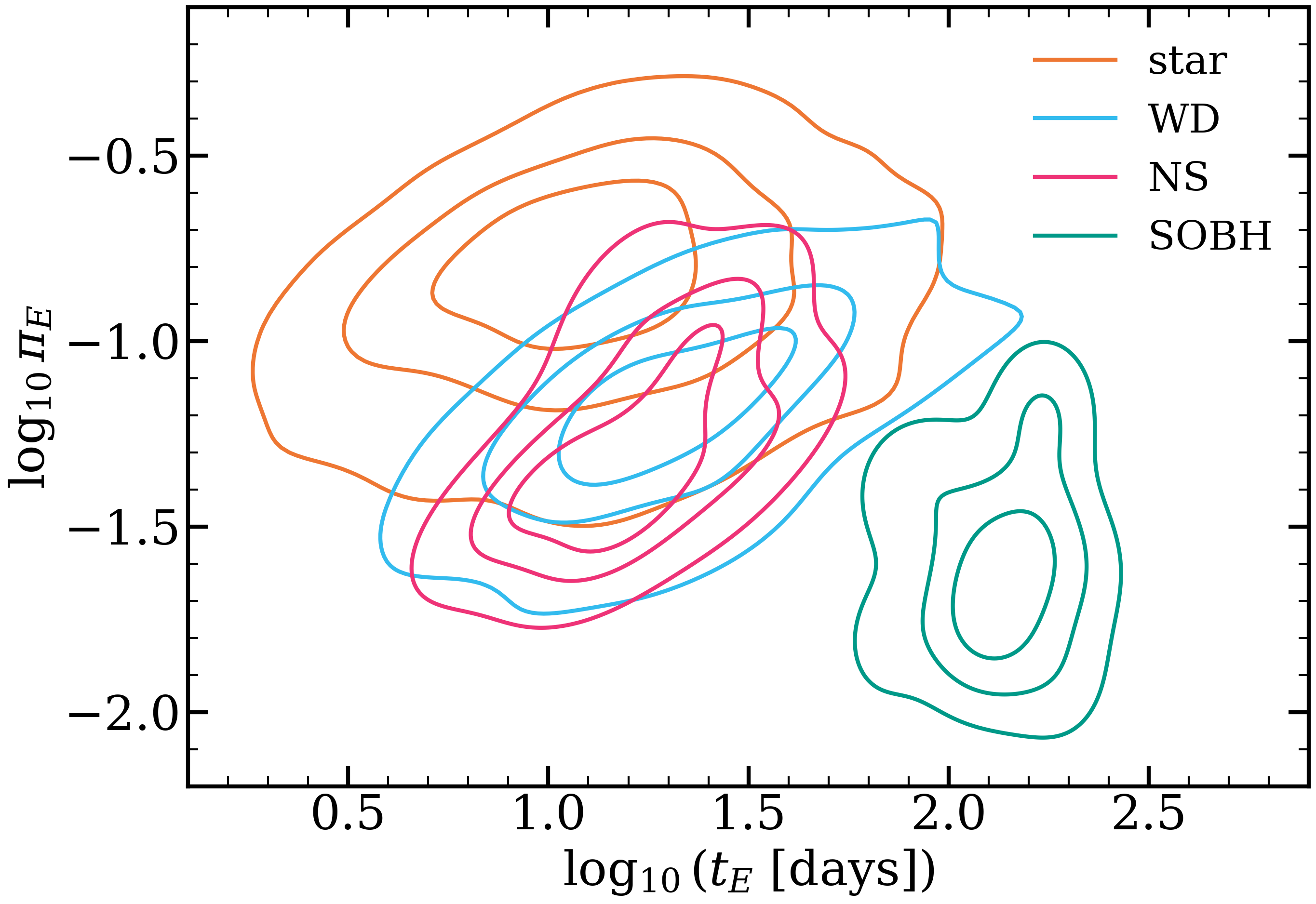}
    \caption{Simulation of microlensing events using the Galactic model parameters in Table \ref{tab:simulation} and the \cite{Sukhbold2016} IFMR projected into $\log_{10} t_{\rm E}$ -- $\log_{10} \pi_{\rm E}$ space. Events with different lens classes occupy different but overlapping regions of this space. The contours enclose 30\%, 60\% and 90\% of the probability mass for each class and were calculated using KDEs of the simulated catalog of events.}
    \label{fig:kdes}
\end{figure}

From the Galactic simulation, we have a simulated catalog of events labeled by lens type with predicted distributions in the space of microlensing light curve parameters. 
Fig. \ref{fig:kdes} shows the simulated catalog of events in the $t_{\rm E}-\pi_{\rm E}$ space separated by lens class: star, WD, NS or SOBH. 
We can see that events with different lens classes lie in different but overlapping regions of this space making it useful for classifying events. 
In Fig. \ref{fig:kdes} and the work that follows, we use a kernel density estimate (KDE) via \texttt{scipy} \citep{Virtanen2020} with a bandwidth determined using Scott's rule \citep{scott_bandwidth} to calculate the probability density functions for each lens class in all microlensing parameters. 

Following \citet{Perkins2024}, we can leverage these simulation predictions from the model of the Galaxy to classify the lens of a given event. We can calculate the probability that a microlensing event has a particular class given its lightcurve data, $\mathbf{d}$, and the Galactic model, $\mathcal{G}$,
\begin{equation}
p(\text{class}_L| \boldsymbol{d}, \mathcal{G}) = \frac{p(\text{class}_L| \mathcal{G})p(\boldsymbol{d}| \text{class}_L, \mathcal{G})}{p(\boldsymbol{d}| \mathcal{G})}.
\label{eq:bayes}
\end{equation}
where $\text{class}_{L}\in\{\text{Star}, \text{WD}, \text{NS}, \text{SOBH} \}$ denotes the lens class. We can now write Eq. (\ref{eq:bayes}) in a form that can be computed by introducing parameters of the microlensing light curve $\phi=\left[t_{\rm E}, \pi_{\rm E},...\right]$,
\begin{equation}
    p(\text{class}_L | \mathbf{d}, \mathcal{G}) =
    \frac{p(\text{class}_L | \mathcal{G})}{p(\mathbf{d} | \mathcal{G})} \int p(\mathbf{d} | \mathbf{\phi}) p(\mathbf{\phi} | \text{class}_L, \mathcal{G}) d\mathbf{\phi}.
    \label{eq:class_eq}
\end{equation}
The integral can be then approximated via an importance sampling approximation using $S$ independent posterior samples, ${\phi}_c$, obtained by fitting the lightcurve, $\mathbf{d}$, under some prior $\mathbf{\pi}(\mathbf{\phi})$,
\begin{equation}
\label{eq:sampling}
    \int p(\mathbf{d} | \mathbf{\phi}) p(\mathbf{\phi} | \text{class}_L, \mathcal{G}) d\mathbf{\phi}\approx\frac{1}{S}\sum_{c=0}^{S}\frac{p(\mathbf{\phi}_c | \text{class}_L, \mathcal{G})}{\mathbf{\pi}(\mathbf{\phi}_c)}.
\end{equation}
Here, $p(\mathbf{\phi}_c | \text{class}_L, \mathcal{G})$ is calculated by using the Galactic model class KDE estimate seen in Fig. \ref{fig:kdes} and $\mathbf{\pi}(\mathbf{\phi})$ must have support over the class KDE for the importance sampling approximation to be reliable.

\begin{figure*}[t!]
    \centering
    \includegraphics[width=\textwidth, trim=0cm 0cm 0cm 0cm]{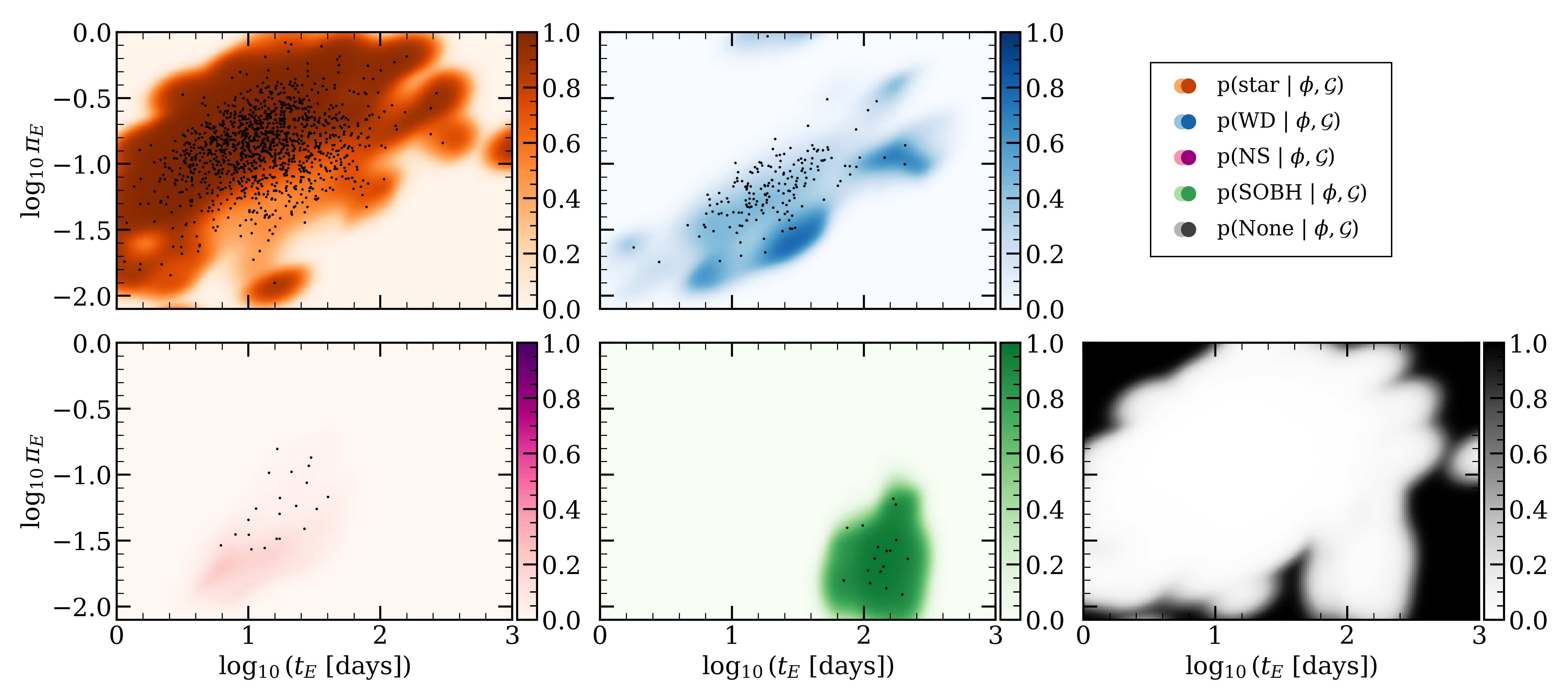}
    \caption{Relative lens classification probabilities in $\log_{10} t_{\rm E}$ -- $\log_{10} \pi_{\rm E}$ space for the \texttt{PoPSyCLE} model detailed in Table \ref{tab:simulation} with the \cite{Sukhbold2016} initial final mass relation, evaluated on a 1000x1000 grid. $p(\text{class}_L | \mathbf{\phi}, \mathcal{G})$ for each gridpoint $\mathbf{\phi}$ are calculated and normalised to 1 as in Eq. \ref{eq:bayes} (with the evaluated distribution $\boldsymbol{d}$ reduced to a single point $\mathbf{\phi}$) -- i.e., the colors represent probabilities of an event being classified as each class, if its parameters $\mathbf{\phi}$ were exactly known. \\ The black points in each panel are simulated events belonging to the corresponding classes. The None class is constructed to have probability density in regions of low or zero Galactic model support. The WD and NS panels show that there are only relatively small or no regions which an event can have a high probability of being classified as such. In contrast, the Star and SOBH panels show regions in the space where high-confidence lens classifications can be made.}
    \label{fig:kdespace}
\end{figure*}

In Eq. (\ref{eq:class_eq}), $p(\text{class}_L | \mathcal{G})$ is the fraction of events that have class$_L$ in the Galactic model, $\mathcal{G}$. Assuming that our set of considered lens classes is complete, the evidence of a single lens (the denominator of Eq. \ref{eq:class_eq}) is,
\begin{equation}
    p(\boldsymbol{d} | \mathcal{G}) = \sum_{\text{class}_L\in\text{classes}} p(\text{class}_L|\mathcal{G}) p(\boldsymbol{d}|\text{class}_L, \mathcal{G}),
\end{equation}
which normalizes the class probabilities to unity. In this work, we chose to link $\mathcal{G}$ to the lens classification using the subset of microlensing lightcurve  parameters $\phi \ {\equiv} \ [\log_{10} t_{\rm E}, \ \log_{10} \pi_{\rm E}]$ which has been demonstrated to be an effective space for delineating lens classes \citep[e.g.,][]{Lam2020, golovich, Perkins2024, Fardeen2024, Pruett2024}. Choosing a subset of the events' parameters to link to $\mathcal{G}$ to lens class is an approximation and this classifier could in principle be used with any subset of parameters or the complete set. In this work we choose $\phi \ {\equiv} \ [\log_{10} t_{\rm E}, \ \log_{10} \pi_{\rm E}]$ because this space shows intrinsic separation in lens class, can be constrained from the event lightcurve, and KDEs are relatively robust and fast to construct in two dimensions. 

\subsection{Epistemic uncertainty of the Galactic model}

Galactic models that produce Monte Carlo simulations of microlensing events can be incomplete in two main ways. Firstly, they can simply be incomplete in their lens populations. For example, the \texttt{PopSyCLE} simulation used in this work does not contain sub-stellar lenses below $0.07M_{\odot}$ \citep{Lam2020} such as brown dwarfs or free-floating planets \citep[e.g.,][]{Johnson2020}. Secondly, Monte Carlo simulations can be noisy and incomplete in the tails of the event parameter distributions especially given the computation cost of microlensing simulations.

Over the course of developing the classifier, we encountered issues with the noisy tails of the simulation class KDEs in the $\log_{10} t_{\rm E} - \log_{10} \pi_{\rm E}$ space, related to the second aforementioned problem. For example, despite the WD and NS distributions being nested in the wings of a far more numerous and dominant Star distribution (see Fig. \ref{fig:kdes}) -- so the probability of a WD or NS lens should be $\ll 1.0$ -- the classifier predicted $p(\text{class}_L | \mathbf{d}, \mathcal{G})\sim 1.0$ for $\text{class}_L \in$ (WD, NS) for some events. In these cases, the events in question had posterior distributions with many samples in regions of $\log_{10} t_{\rm E} - \log_{10} \pi_{\rm E}$ that had little or no simulation support from the Galactic simulation. In turn, this caused the KDE estimates of the WD and NS classes to be extrapolated and evaluated in their noisy tails well beyond any simulation support.

To remedy this issue, we introduced a additional None class that has non-zero density in regions of $\log_{10} t_{\rm E} - \log_{10} \pi_{\rm E}$ that do not have any samples from the Galactic model for any class. To construct the probability density for this class we define a grid of 1000x1000 bins over the entire $\log_{10} t_{\rm E}$ -- $\log_{10} \pi_{\rm E}$ subspace allowed by the priors used in event modelling ($0.5 \text{ days} < t_{\rm E} < 3000 \text{ days}$, $10^{-5} < \pi_{\rm E} < 3$). We evaluate the density of a two-dimensional KDE constructed using positions of all simulated events, $p(\mathbf{\phi} | \mathcal{G})$. We use a tophat KDE implemented in \texttt{scikit-learn} \citep{Pedregosa2011} for a sharp slope between covered and uncovered regions, with a bandwidth of $0.4$ chosen to be minimal while still returning a simply connected space of non-zero values over the bulk of $\log_{10} t_{\rm E}$ -- $\log_{10} \pi_{\rm E}$ space. We then assign values $p(\phi | \text{None})$ to bin centers, where: 
\begin{equation}
    p(\phi | \text{None},\mathcal{G}) = A \left( 1 - \frac{p(\mathbf{\phi} | \mathcal{G})}{\max_{\phi} (p(\mathbf{\phi} | \mathcal{G}))} \right)
    \label{eq:none}
\end{equation} 
and $A$ is a constant that normalizes probability density function to unity over the entire subspace. We assign the prior probability of the None class $p(\text{None})=0.01$. This class is then included as one of the classes in the set of classes in Section \ref{sec:bayes}. Fig. \ref{fig:kdespace} shows the relative class probabilities in $\log_{10} t_{\rm E} - \log_{10} \pi_{\rm E}$ space including the None class. Comparing Figs. \ref{fig:kdes} and \ref{fig:kdespace} shows that the None class is highest in regions of low and zero Galactic model support. Fig. \ref{fig:kdespace_old} shows the effect of the adding the None class on the tails of the WD class probility distribution. 

This None class can also partially mitigate for Galactic model that has an incomplete lens population such as \texttt{PopSyCLE}. In the case of missing brown dwarfs and free-floating planet lenses, which can have largely non-overlapping distributions in $\log_{10} t_{\rm E} - \log_{10} \pi_{\rm E}$ to the stellar distributions (i.e., $\log_{10} t_{\rm E} < 0.5$), the None class will absorb event posterior density in those regions and the classifier will return a large None class probability. 

Overall, the None class described in this section traces the amount of posterior density an event has in region of low or zero Galactic population support. Events with high None class probabilities should be treated with caution and generally investigated further.

For additional lens classes that are not included in the Galactic model, but that overlap with the bulk of the simulation distribution (e.g., binary systems; Abrams, et. al. in review), the None class is not effective as it has very small probability density in these regions by definition (see Eq. \ref{eq:none}). To mitigate against this type of Galactic model uncertainty the resulting classification from many different Galactic models can be compared or averaged over. In this paper we focus on different IFMRs which are likely to affect black hole classification probabilities. We make this classification framework available via the {\tt popclass} python package\footnote{\url{https://github.com/LLNL/popclass}}.

\subsection{Astrometric posterior predictive distributions}\label{sec:thetaE_inference}

In addition to lens classification, this framework can be used to derive posterior predictive distributions. In this work we focus on photometric events and their subsequent astrometric follow-up in the context of stellar-origin black holes. For high-probability SOBH candidates, it is useful to be able to predict whether an event is likely to have a detectable astrometric signal if the lens is indeed a SOBH, so expensive astrometric follow-up observations can be allocated efficiently. This motivates predicting the angular scale of the event $\theta_{\rm E}$. This extension could be also used in the future for any other posterior predictive distributions of interest, such as lens masses or relative proper motions (allowing for selections of events where the luminous lens can be resolved from the source).

We can take the posterior constraints on the \emph{photometric} microlensing parameters (e.g., $t_{\rm E}$ and $\pi_{\rm E}$) and, using the Galactic model, predict distributions on the parameters consistent with, but not directly constrained by the light curve such as $\theta_{\rm E}$. The posterior predictive density on $\theta_{\rm E}$ is then directly related to the predicted maximum possible astrometric deviation $\delta_{ \rm max}$, which occurs at lens-source separation of $\sqrt{2}\theta_{\rm E}$ \citep{Dominik2000}:
\begin{equation}
    \delta_{ \rm max} = \frac{\sqrt{2}}{4} \theta_{\rm E}.
\end{equation}
The posterior predictive distribution on $\theta_{\rm E}$ given light curve data, $\textbf{d}$, and Galactic model, $\mathcal{G}$, is,
\begin{align}\nonumber
    &p(\theta_{\rm E} | \mathcal{G}, \textbf{d}, \text{class}_{L}=\text{SOBH}) \\\nonumber
    &= \int p(\theta_{\rm E}, \phi | \mathcal{G}, \textbf{d}, \text{class}_{L}=\text{SOBH}) d\phi\,, \\
    &= \int \frac{p (\textbf{d} |\phi)p(\theta_{\rm E}, \phi | \mathcal{G}, \text{class}_{L}=\text{SOBH})}{p(\textbf{d}|\mathcal{G},\text{class}_{L}=\text{SOBH})} d\phi\,,
    \label{eq:thetaE_pred}
\end{align}
where $\phi$ represents the photometric microlensing parameters. 
Note that the likelihood $p(\textbf{d}| \phi)$ is independent of $\theta_{\rm E}$, as $\textbf{d}$ represents the \emph{photometric} data.
This calculation compares the event's photometric likelihood to the expected distribution from a population model. While the mapping is not fully specified to translate the photometric parameters to $\theta_{\rm E}$, enfolding restrictions from the population model restricts the $\theta_{\rm E}$ space which is still consistent with the photometric likelihood.

To sample the posterior predictive distribution for a high probability SOBH candidate, we estimate $p(\theta_{\rm E}, \phi | \mathcal{G}, \text{class}_{L}=\text{SOBH})$ by building a three-dimensional KDE over the Galactic simulation samples in the $\log_{10} t_{\rm E} - \log_{10} \pi_{\rm E} -\theta_{\rm E}$ space. Assuming our golden-sample candidates are in fact black holes, we now set out to assess their observed signal and chances for detection. We infer the $\theta_{\rm E}$ distribution for each candidate, using a three-dimensional $\log_{10} t_{\rm E} - \log_{10} \pi_{\rm E} - \theta_{\rm E}$ KDE constructed on all simulated SOBH events from the Sukhbold N20 run and following Sec.~\ref{sec:thetaE_inference}. While there is no prior in Eq.~\eqref{eq:thetaE_pred}, as this is a predictive distribution and not a posterior distribution, we define a uniform prior in $\theta_E$ because {\tt dynesty} requires one and a uniform prior does not alter the results. We define a $\mathcal{U}(-2, 15 \ [\textnormal{mas}])$ prior for $\theta_{\rm E}$, allowing for a small margin of unphysical negative $\theta_{\rm E}$ to capture noisy distributions around near-zero values. We use the dynamic nested sampling algorithm \citep{Higson2019} implemented by \citet{DYNESTY} in the {\tt dynesty} code. We use random walk sampling \citep{Skilling2006} with multiple bounding ellipsoids and $1\,000$ initial live points. We allocate samples with 100\% of the weight placed on the posterior and use the default stopping function.

\begin{figure*}[t!]
    \centering
    \includegraphics[width=\textwidth, trim=4cm 0cm 0cm 1cm]{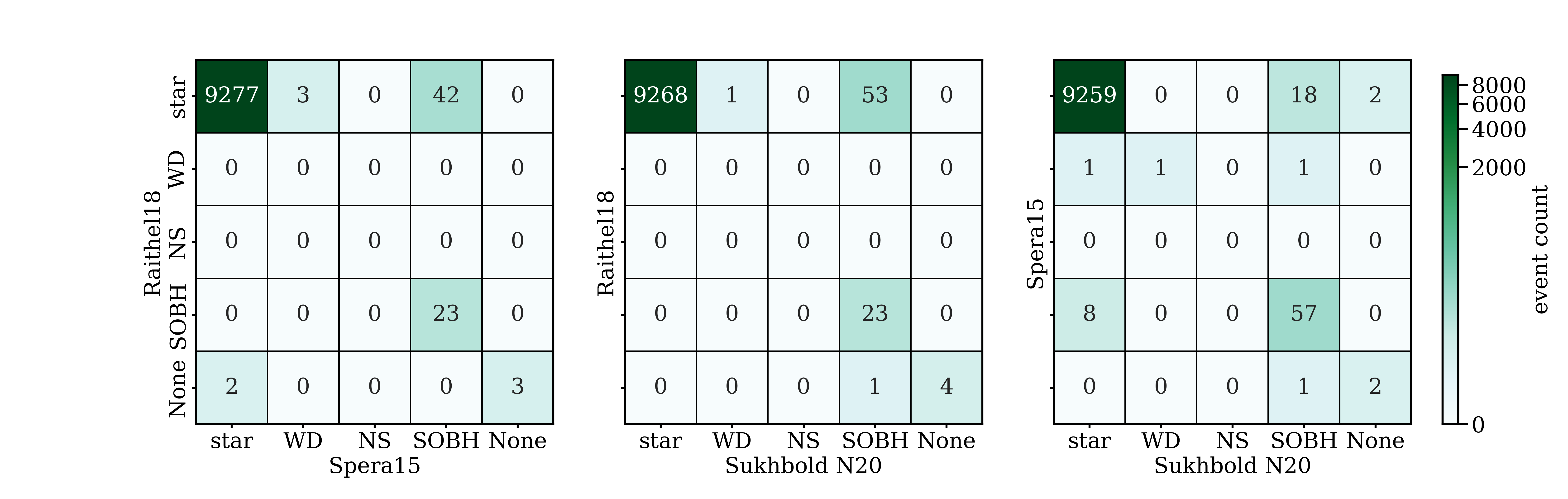}
    \caption{Pairwise confusion matrices between classifications based on three different Galactic model differing only in their initial-final mass relations (IFMRs): Sukhbold N20 \citep[based on models from][]{Sukhbold2014, Sukhbold2016, Woosley2017, Woosley2020}, Spera15 \citep{Spera2015}, and Raithel18 \citep{Raithel2018}. For a given Galactic model, the lens class is assigned to be the class with the highest posterior probability. Events appearing along the diagonal means the two different Galactic models agree on the lens classification. All IFMRs agree on bulk of the events being classified as Stars. The right panel shows Sukhbold N20 and Spera15 agree on the most SOBH lens classifications (57). The left and middle panels show that Raithel18 classifies most (42/53) events as Stars that were assigned to SOBH candidates in the other (Spera15/Sukhbold N20) IFMRs. The 23 events on which Raithel18 agrees with the other IFMR are the same in the left and middle panels and constitute our golden sample of SOBH candidates. See Section \ref{sec:candidates} for discussion.}
    \label{fig:confusion_ifmrs}
\end{figure*}

\begin{figure}[h!]
    \centering
    \includegraphics[width=\columnwidth]{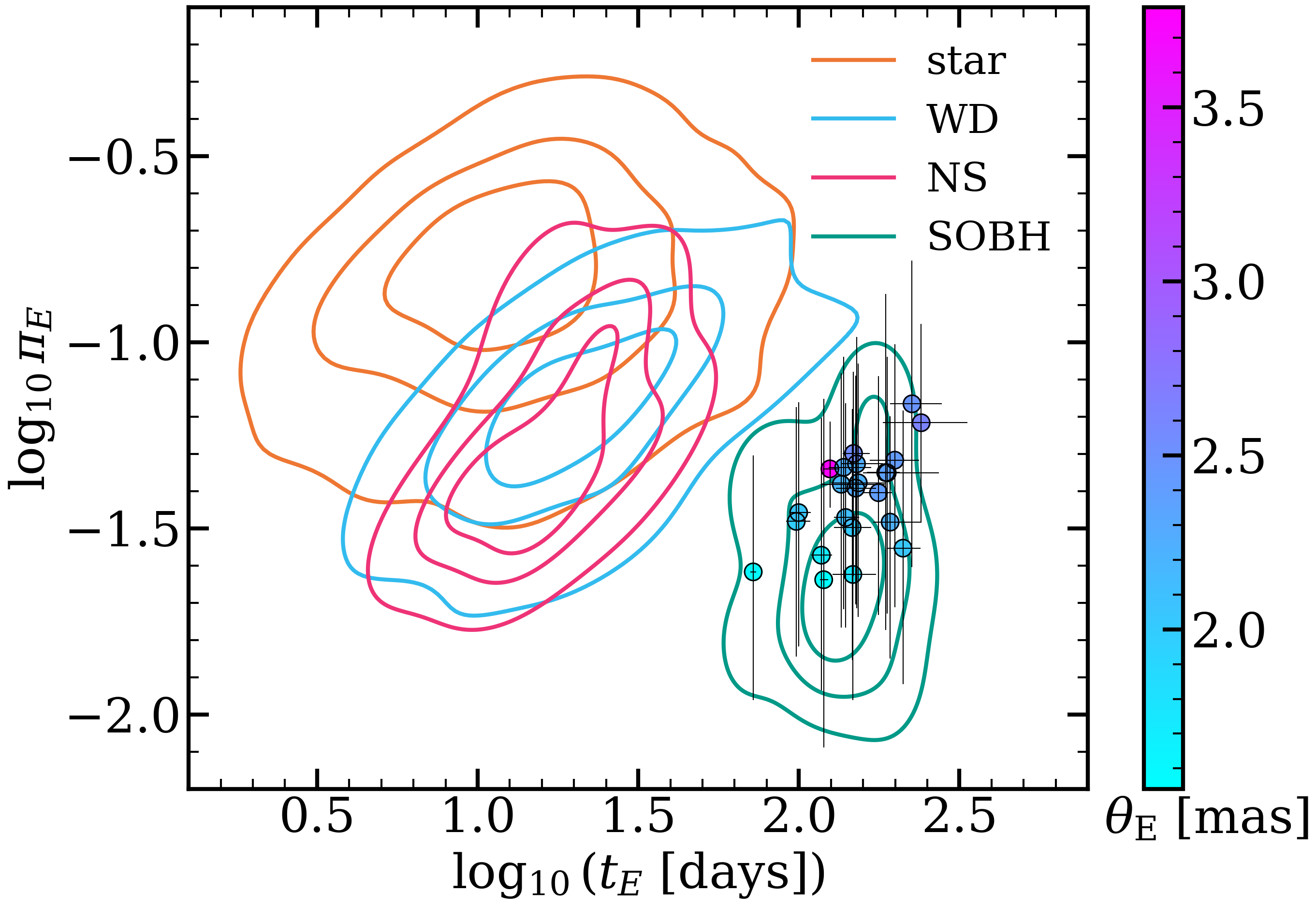}
    \caption{The 23 black hole candidates found in this work presented in $\log_{10} t_{\rm E}$ -- $\log_{10} \pi_{\rm E}$ space. Errorbars indicate 16-84 percentiles of all posterior samples. Colours correspond to median $\theta_{\rm E}$ from the nested sampling posteriors. Iso-contours (as in Figure~\ref{fig:kdes}) represent KDEs of all astrophysical classes in the simulation output using the \citet{Sukhbold2016} IFMR.}
    \label{fig:bh_candidates}
\end{figure}

To compute the integral in Eq. (\ref{eq:thetaE_pred}), we use the posterior distribution samples from the photometric analysis and the importance-sampling technique described in Section \ref{sec:classification} which gives,
\begin{align}\nonumber
    &p(\theta_{\rm E} | \mathcal{G} , d, \text{class}_{L}=\text{SOBH}) \\
    &\propto \frac{1}{S} \sum_i^S \frac{p(\theta_{\rm E}, \hat{\phi}_i | \mathcal{G}, \text{class}_{L}=\text{SOBH}) }{\pi(\hat{\phi}_i)} \,,\\
    \hat{\phi}_i&\sim p(\phi_i | d) \,.
\end{align}
$\pi(\phi)$ denotes the prior density function used in modelling the events; here, for $\log_{10} t_{\rm E}$ and $\log_{10} \pi_{\rm E}$ it was a truncated normal distribution that can be found in Table 1 of \cite{golovich}.

\section{Classification results}
\label{sec:classification_results}

\subsection{High-probability black hole candidates}\label{sec:candidates}

We classify the event sample with three different Galactic models only differing in their underlying initial-final mass relations: Sukhbold N20 \citep[based on models from][]{Sukhbold2014, Sukhbold2016, Woosley2017, Woosley2020}, Spera15 \citep{Spera2015}, and Raithel18 \citep{Raithel2018}. A detailed description of all IFMRs and their integration into the \texttt{PopSyCLE} simulation is provided in \citet{Rose2022}.
For each event and each IFMR, we assign a class corresponding to the highest $p(\text{class}_a | \mathbf{d}, \mathcal{G})$ value. We summarise those classifications in confusion matrices between pairs of IFMRs in Fig.~\ref{fig:confusion_ifmrs}. 

We find that while the three different IFMRs relations agree on classifying the majority of lenses as stars, the SOBH classifications are impacted by the underlying IFMR. The Raithel18 models yields significantly less black hole candidates than the other two, only finding 23 common SOBHs with Sukhbold N20 and Spera15. This is in contrast to Spera15 and Sukhbold N20 which agree on 57 SOBH candidates. 

The reason for this might be two-fold. Firstly, the Raithel18 simulation simply generates fewer black hole lenses; the prior black hole lens probability is equal to 0.0083 (as compared to 0.0100 and 0.0116 with the Spera15 and Sukhbold N20 simulation output, respectively). Secondly, the simulated Raithel18 black hole lenses are significantly less massive (on average 9.3$M_{\odot}$, compared to 12.6$M_{\odot}$ for Spera15 and 12.2$M_{\odot}$ for Sukhbold N20); this may cause the population to be less well-separated from non-SOBHs in $\log_{10} t_{\rm E} - \log_{10} \pi_{\rm E}$ space.
Our observations reflect those of \citet{Rose2022}, who note that the Raithel18 IFMR only produces black holes in the 5-16$M_{\odot}$ range, while the other two include significantly higher masses. \citet{Rose2022} attribute the difference to the Raithel18 IFMR assuming solar metallicity for all progenitors, hence missing the most massive stellar remnants formed from the low-metallicity population.

We define the best black hole candidates as events that were assigned the SOBH class in classifications based on all three IFMRs. We give an overview of the 23 found candidates in Figure~\ref{fig:bh_candidates} and Table \ref{tab:bh_candidates_overview}. This table is a subset of rows and columns from the full classification result table, where we report probabilities for each class under each IFMR for each event; due to the size of the full table, we will publish it as an auxiliary dataset upon acceptance of this paper.

\begin{table*}[]
\centering

\begin{tabular}{p{2.9 cm} p{1.65cm} p{1.85 cm} p{1.9 cm} p{1.9 cm} p{2.1 cm} cp{2.5 cm}}
\hline \hline
OGLE ID & $\alpha$ [$^\circ$] & $\delta$ [$^\circ$] & $\log_{10} (t_{\rm E} \text{[d]})$ & $\log_{10} \pi_{\rm E}$ & $b_{ \rm sff}$ & $p(\text{SOBH} | \mathbf{d}, \mathcal{G}_{\rm SN20})$ \\
\hline
BLG117.1.89360 & 267.51625 & -35.61075 & $2.166^{+0.058}_{-0.055}$ & $-1.50^{+0.32}_{-0.35}$ & $0.036^{+0.006}_{-0.005}$ & 0.864 \\
BLG131.1.104549 & 267.41833 & -34.43564 & $2.132^{+0.134}_{-0.062}$ & $-1.38^{+0.30}_{-0.38}$ & $0.614^{+0.336}_{-0.368}$ & 0.793 \\
BLG195.1.373 & 268.42996 & -29.52431 & $2.180^{+0.081}_{-0.046}$ & $-1.33^{+0.34}_{-0.39}$ & $0.719^{+0.258}_{-0.348}$ & 0.797 \\
BLG196.5.68751 & 270.84188 & -29.25872 & $2.077^{+0.012}_{-0.010}$ & $-1.64^{+0.48}_{-0.45}$ & $0.566^{+0.025}_{-0.028}$ & 0.788 \\
BLG208.3.222797 & 271.59067 & -28.74619 & $1.858^{+0.008}_{-0.008}$ & $-1.62^{+0.31}_{-0.34}$ & $0.590^{+0.013}_{-0.013}$ & 0.428 \\
BLG216.2.201174 & 270.74992 & -28.29794 & $1.992^{+0.042}_{-0.030}$ & $-1.48^{+0.31}_{-0.36}$ & $0.771^{+0.171}_{-0.181}$ & 0.641 \\
BLG233.3.73254 & 270.74750 & -27.06481 & $2.177^{+0.062}_{-0.060}$ & $-1.39^{+0.30}_{-0.31}$ & $0.474^{+0.178}_{-0.119}$ & 0.860 \\
BLG503.16.64872 & 267.16775 & -34.85408 & $2.299^{+0.074}_{-0.076}$ & $-1.32^{+0.31}_{-0.39}$ & $0.023^{+0.005}_{-0.005}$ & 0.767 \\
BLG507.31.122188 & 269.01404 & -31.13247 & $2.097^{+0.027}_{-0.021}$ & $-1.34^{+0.13}_{-0.10}$ & $0.851^{+0.142}_{-0.157}$ & 0.793 \\
BLG515.15.27802 & 270.33738 & -32.36358 & $2.284^{+0.097}_{-0.052}$ & $-1.48^{+0.28}_{-0.36}$ & $1.399^{+0.567}_{-0.620}$ & 0.888 \\
BLG518.30.17228 & 271.98604 &-26.30169 & $2.325^{+0.053}_{-0.050}$ & $-1.55^{+0.37}_{-0.36}$ & $0.627^{+0.126}_{-0.109}$ & 0.837 \\
BLG520.10.64977 & 272.51396 & -29.35044 & $2.270^{+0.042}_{-0.057}$ & $-1.35^{+0.48}_{-0.42}$ & $0.207^{+0.069}_{-0.039}$ & 0.748 \\
BLG580.12.77934 & 272.12962 & -25.60019 & $2.185^{+0.085}_{-0.082}$ & $-1.38^{+0.32}_{-0.36}$ & $0.347^{+0.130}_{-0.091}$ & 0.830 \\
BLG605.20.76630 & 266.52504 & -36.58792 & $2.352^{+0.091}_{-0.067}$ & $-1.17^{+0.38}_{-0.44}$ & $1.484^{+0.542}_{-0.688}$ & 0.545 \\
BLG632.18.118008 & 266.89000 & -23.18578 & $2.140^{+0.084}_{-0.046}$ & $-1.34^{+0.30}_{-0.38}$ & $1.332^{+0.453}_{-0.575}$ & 0.797 \\
BLG632.20.147708 & 266.50442 & -23.33761 & $2.168^{+0.070}_{-0.061}$ & $-1.62^{+0.28}_{-0.34}$ & $0.588^{+0.281}_{-0.192}$ & 0.911 \\
BLG633.25.85822 & 265.70662 & -24.61531 & $2.000^{+0.036}_{-0.028}$ & $-1.46^{+0.30}_{-0.36}$ & $0.943^{+0.189}_{-0.198}$ & 0.649 \\
BLG638.18.94340 & 268.12804 & -22.64825 & $2.381^{+0.142}_{-0.118}$ & $-1.22^{+0.26}_{-0.27}$ & $0.161^{+0.104}_{-0.067}$ & 0.577 \\
BLG643.28.35495 & 269.15404 & -22.90167 & $2.170^{+0.050}_{-0.032}$ & $-1.30^{+0.22}_{-0.33}$ & $1.047^{+0.200}_{-0.294}$ & 0.841 \\
BLG645.26.75287 & 269.83983 & -26.15864 & $2.070^{+0.031}_{-0.029}$ & $-1.57^{+0.37}_{-0.36}$ & $0.749^{+0.172}_{-0.135}$ & 0.846 \\
BLG652.18.94827 & 265.78996 & -25.86931 & $2.275^{+0.159}_{-0.074}$ & $-1.35^{+0.31}_{-0.38}$ & $0.829^{+0.550}_{-0.518}$ & 0.794 \\
BLG661.12.38593 & 264.95212 & -33.81094 & $2.248^{+0.040}_{-0.046}$ & $-1.40^{+0.31}_{-0.33}$ & $0.129^{+0.034}_{-0.021}$ & 0.878 \\
BLG662.21.34275 & 262.55525 & -30.24875 & $2.146^{+0.041}_{-0.034}$ & $-1.47^{+0.31}_{-0.29}$ & $1.077^{+0.226}_{-0.206}$ & 0.906 \\
\hline
\end{tabular}
\caption{Overview of the 23 black hole candidates. We present OGLE IDs (field number + star number) and coordinates, posterior distribution statistics for timescale, parallax and blending from the modelling of \citet{golovich}, and the classifier-reported probability of belonging to the SOBH class under the Sukhbold N20 IFMR. $\log_{10} (t_{\rm E} \text{[d]})$, $\log_{10} \pi_{\rm E}$ and $b_{ \rm sff}$ distributions are represented as median values with 16-84 percentile intervals. This table is a subset of the full classification results; probabilities for all classes, all IFMRs and all events will be made available in the online supplementary material.}
\label{tab:bh_candidates_overview}
\end{table*}

\begin{figure}[t!]
    \centering
    \includegraphics[width=\columnwidth]
    {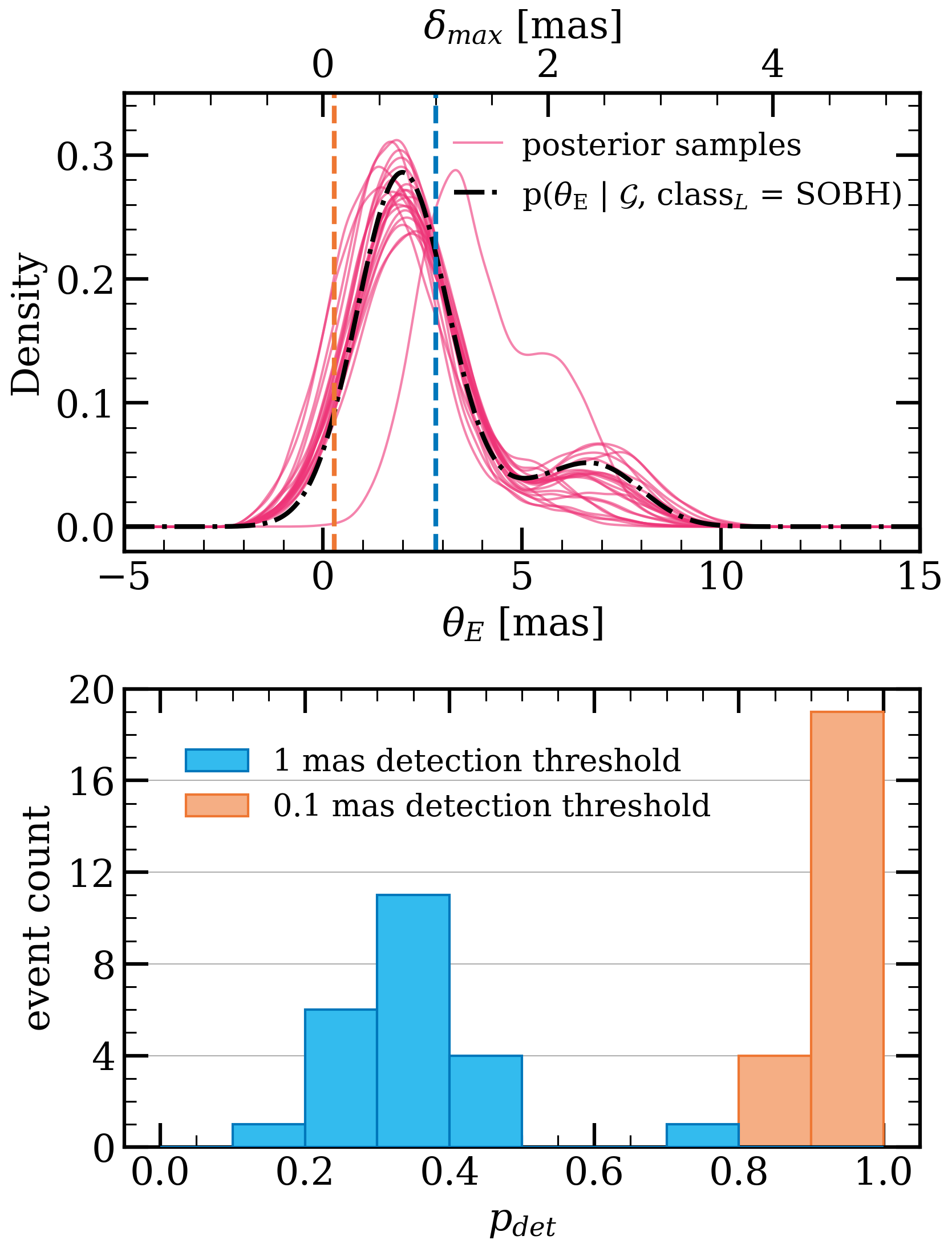}
    \caption{{\it Top:} Posterior predictive $\theta_{\rm E}$ distributions with corresponding maximum astrometric deviation $\delta_{\rm max}$ from nested sampling for all 23 strong SOBH candidates (represented as 10000 posterior samples smoothed with a 1D Gaussian KDE). Vertical dashed blue line represents the precision of {\it Hubble Space Telescope}-like astrometric follow-up ($\sigma_{\rm obs} = $ 1 mas), while the respective orange line represents that of the upcoming {\it Roman Space Telescope} ($\sigma_{\rm obs} = $ 0.1 mas). Dashed-dotted black line represents the prior predictive distribution $p(\theta_{\rm E} | \mathcal{G}, \text{class}_L = \text{SOBH})$ constructed from the simulated events. {\it Bottom:} A histogram of probability of detection of astrometric signal from the candidates given a 1 mas (blue) or 0.1 mas (orange) detection threshold, allocated at optimal times (including peak astrometric signal and after the event), estimated as $p_{ \rm det} = p(\delta_{\rm max} > \sigma_{\rm obs})$ and integrated over all posterior samples for each event.
    }
    \label{fig:thetaE_distr}
\end{figure}

We find that the inferred $\theta_{\rm E}$ probability distribution is bimodal for all candidates. We present all $\theta_{\rm E}$ distributions in Fig.~\ref{fig:thetaE_distr}. We estimate the probability that the astrometric signal of an event could be detected by {\it Hubble Space Telescope}- or {\it Roman Space Telescope}-like follow-up, assuming a criterion of $\delta_{\rm max} > 1 \ \textnormal{mas}$ and $> 0.1 \ \textnormal{mas}$ respectively and checking the fraction of $\theta_{\rm E}$ samples for the event passing this criterion.

\begin{figure}[t!]
    \centering
    \includegraphics[width=\columnwidth, trim=0cm 0cm 0cm 0cm]{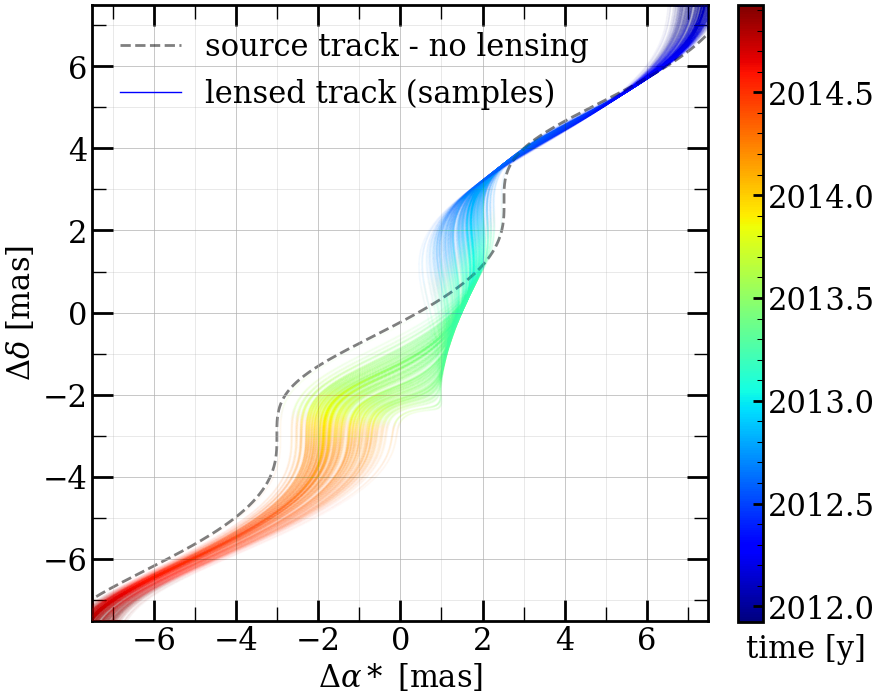}
    \caption{Simulated possible astrometric tracks for black hole candidate BLG507.31.122188, which would be the most likely to have detectable astrometric signal if it had been allocated follow-up observations. Solid lines represent 300 randomly drawn $\theta_{\rm E}$ values from posterior samples; the dotted line represents the source track. Color coding represents time (from 1.5 years before to 1.5 years after the event maximum). The scale is centered so that the straight-line source motion is at (0, 0) during the event maximum. 1 milliarcsecond -- the fine gridpoint separation -- corresponds to typical astrometric precision of {\it Hubble Space Telescope} observations, or along-scan precision of \gaia~observations at $G\approx$ 18 mag (event brightness out of amplification).}
    \label{fig:astrom_signal_507}
\end{figure}

To illustrate the expected signal, we simulate astrometric tracks for the best candidate, BLG507.31.122188, which has the highest probability of detectable signal -- 77\% -- and $b_{ \rm sff}$ (the ratio of flux coming from the source to total observed flux) of $0.851^{+0.142}_{-0.157}$. Due to the source/total flux ratio being consistent with 1 (at 1.05$\sigma$), and keeping in mind the higher spatial resolution of \gaia~compared to OGLE, we assume this event to be unblended in \gaia. We match this candidate with a \gaia~source of magnitude $G = 18.00$ and an available 5-parameter astrometric solution; we assume \gaia~observations should not be highly impacted by lensing signal, as the event peaked in mid-2013 with a $t_{\rm E}$ of 125 days. We fix the photometric lensing parameters ($t_0, u_0, t_{\rm E}, \pi_{\rm E}, \phi$) at their median values from modelling; we fix the astrometric motion of the source ($\mu_{\alpha*,S}, \mu_{\delta,S}, \pi_S$) at values reported by \gaia; finally, we sample the remaining parameter $\theta_{\rm E}$ from the posterior predictive distribution. We draw 300 $\theta_{\rm E}$ values from posterior samples and use the {\tt astromet}\footnote{\url{https://github.com/zpenoyre/astromet.py}} package to plot the simulated tracks in Figure~\ref{fig:astrom_signal_507}.

As the events analysed in this work happened between 2002 and 2017, it is not possible to schedule follow-up for an independent astrometric $\theta_{\rm E}$ measurement. However, for some of the OGLE-IV events (spanning 2010-2017) there might be archival astrometric \gaia~data. The \gaia~mission \citep{gaia} has been collecting data since July 2014 and is scheduled to operate until 2025. The next data release, \gaia~DR4, will include astrometric time-series from the first 5.5 years of observations.

\begin{center}
\begin{table*}[t!]
\hspace*{-2.5cm}\begin{tabular}
{ p{2.6cm} p{1.2cm} p{1.1cm} p{0.9cm} p{1.6cm} p{0.9cm} p{1.6cm} p{0.9cm} p{1.8cm} p{1.3cm}}
\hline \hline
OGLE ID & $t_{0,50}$ [year] & RUWE & $\bar{G}_{\text{obs}}$ [mag] & $\sigma_{\rm AL} (\bar{G}_{\text{obs}})$ [mas] & $G_{\uparrow, \text{obs}}$ [mag] & $\sigma_{\rm AL}(G_{\uparrow, \text{obs}})$ [mas] & $G_{\uparrow, \text{pred}}$ [mag] & $\sigma_{\rm AL}(G_{\uparrow, \text{pred}})$ [mas] & $\delta_{\rm{max}}$ [mas]\\ \hline
BLG645.26.75287 & 2014.94 & 1.00 & 18.22 & 1.35 & 17.63 & 0.95 & 16.91 & 0.63 & $0.62^{+0.49}_{-0.44}$ \\ 
BLG662.21.34275 & 2016.80 & 1.18 & 19.73 & 3.95 & 18.47 & 1.61 & 18.36 & 1.49 & $0.75^{+0.53}_{-0.45}$ \\ 
BLG515.15.27802 & 2015.76 & - &	19.82 & 4.21 & - & - & 19.34 & 2.91 & $0.79^{+0.68}_{-0.48}$ \\ 
BLG605.20.76630 & 2014.51 & - &	20.71 &	8.08 & - & - & 20.43 & 6.57 & $0.87^{+1.26}_{-0.52}$ \\ \hline
\end{tabular}
\caption{The sample of 4 SOBH candidate events passing basic cuts for possible \gaia~signal. Median posterior of $t_0$, $t_{0,50}$, given for reference to situate events in the \gaia~mission timeline. RUWE is the astrometric renormalized unit weight error; events with missing RUWE also have missing astrometric solutions. $\bar{G}_{\text{obs}}$ stands for the {\tt phot\_g\_mean\_mag} column in the \gaia~DR3 source catalogue. $\sigma_{\rm AL}(\bar{G}_{\text{obs}})$ is the along-scan astrometric error per observation corresponding to the $\bar{G}_{\text{obs}}$ magnitude.
For sources where DR3 time-series photometry is public (source catalogue variability flag = {\tt VARIABLE}), the brightest observed $G$ value is reported as $G_{\uparrow, \text{obs}}$ and the corresponding astrometric error is reported as $\sigma_{\rm AL}(G_{\uparrow, \text{obs}})$. To estimate expected precision for all events, a lower limit on magnitude and astrometric error is calculated, assuming (optimistically) magnitude at baseline = $\bar{G}_{\text{obs}}$ and magnitude at peak, $G_{\uparrow, \text{pred}}$, corresponding to its amplification at $u = u_{0,50}$ -- the posterior median of $u_{0}$. Events are sorted by increasing $\sigma_{\rm AL}(G_{\uparrow, \text{pred}})$. All $\sigma_{\rm AL}$ values have been calculated with the {\tt astromet} package. $\delta_{\rm{max}}$ is an estimate of the expected astrometric signal -- the median value of the maximum astrometric deviation from posterior $\theta_{\rm E}$ samples with 16-84th percentile intervals. Events with $\delta_{\rm{max}} > \sigma_{\rm AL}$ would be good candidates for detectable signal in time-series astrometry.}
\label{tab:gaia_cm}
\end{table*}
\end{center}

Even before the publication of time-series data, indications from \gaia~astrometric fits can be used to identify astrometric microlensing effects, estimate masses, and determine the nature of the lens \citep{Jablonska2022}. This is possible in the case of events with most prominent astrometric signal (optimally, bright events with high Einstein radii, where the angular scale $\theta_{\rm{E}} \gg$ astrometric measurement error); `bad' fits indicating signal inconsistent with 5-parameter motion can then be exploited to infer the magnitude of the astrometric deviation.

To vet our dataset for possible events where this analysis could be done, we crossmatch the best SOBH candidate sample with the \gaia~DR3 catalogue \citep{gaiaDR3}. Out of the 23 events, 21 have a match (using a 1 arcsec circle around the OGLE coordinates) in \gaia~DR3 data, including 18 with astrometric fit parameters. The incomplete match is likely to be tied to lensing events typically occuring in crowded regions, where \gaia~completeness is significantly lower \citep[e.g.][]{GaiaUnlim}. We find 8 events for which the median $t_0 + t_{\rm E}$ is larger than the start time of \gaia~observations, i.e. they are at least partially covered by \gaia; 7 of them have a \gaia~match.

Astrometry is significantly less likely to contain useful information and more difficult to analyze if events are blended. We use the criterion of $b_{ \rm sff}$ consistent with 1 (within 3$\sigma$, using a Gaussian fit to posterior samples) to further constrain the sample. This is a rough selection cut: blending is rarely well-constrained in the modelling, as it is highly degenerate with other photometric parameters such as $u_0$. Therefore even for the events with $b_{ \rm sff}$ consistent with 1, some solutions with significant blending are allowed. On the other hand, events blended in OGLE do not necessarily have to be blended in \gaia; not only are the observations done in different bands, but most importantly, \gaia's spatial resolution is significantly higher. As anomalous astrometry during amplification can be caused by changing weights in a blended light center of the source and the blend \citep[e.g.,][]{Kaczmarek2022}, and could be misinterpreted as microlensing signal, we choose to keep this criterion as a middle ground. As we make our classifications public, crossmatches with more relaxed or strict criteria can be made in the future.

The blending cut rejects 3 events, which have $b_{ \rm sff}$ with median values of 0.02--0.16 and relatively well-constrained distributions ($0.005 \leq \sigma_{b_{\rm sff}} \leq 0.01$), indicating the blend to be significantly more luminous than the source. We analyze the possible astrometric signal of the remaining 4 candidates and present the results in Table~\ref{tab:gaia_cm}. 

We conclude there is no clear candidate for astrometric signal observable in \gaia. As all 4 found events are relatively faint in the $G$ band, the expected astrometric signal is similar to or lower than the noise level. We also find no indication of excess astrometric signal beyond the 5-parameter fits in the RUWE values. In literature, RUWE values between 1.25 \citep{Penoyre2022} and 1.4 \citep{Lindegren2018, Lindegren2021, Kervella2022} have typically been applied as a cutoff between well-behaved and anomalous (most often binary) sources; \citet{CastroGinard2024} have recently proposed a sky-variable RUWE threshold between 1.15 and 1.37, with the highest values occurring near the Galactic Center. In light of those studies, none of our candidates pass the cut for anomalous motion based on RUWE alone. This does not exclude astrometric signal altogether, but indicates that the astrometric solution is consistent with a 5-parameter single star motion. We suggest the Table~\ref{tab:gaia_cm} events should be revisited in the upcoming \gaia~DR4 time-series data.\\

\subsection{Fast black hole probability approximation}

\begin{figure}[t!]
    \centering
    \includegraphics[width=\columnwidth, trim=0cm 0cm 0cm 0cm]{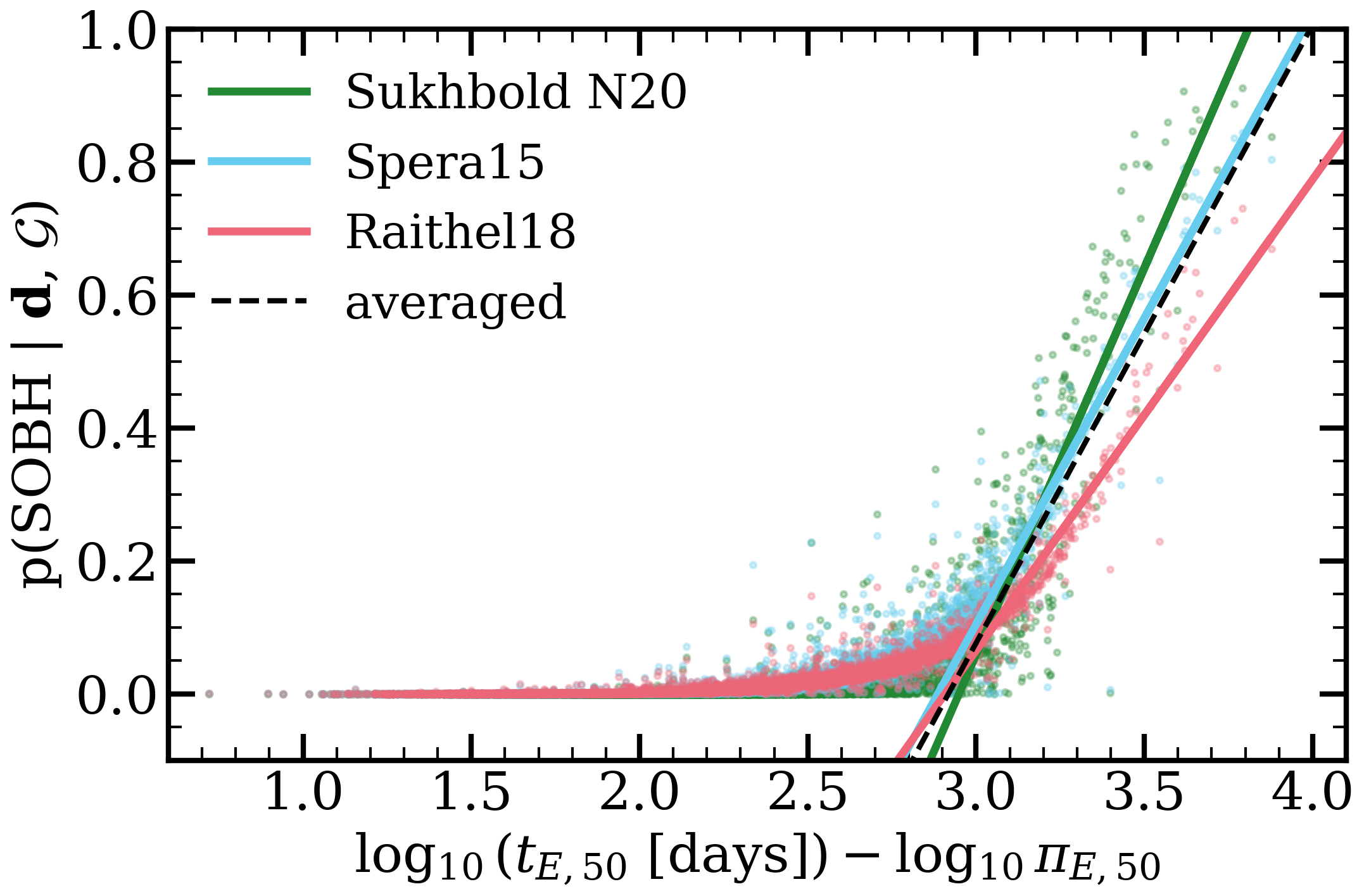}
    \caption{
    Fast approximation of the probability of the lens being a stellar-origin black hole. Colors represent the three IFMRs used. Points represent all OGLE-III and -IV events: the X-axis position is defined as $x = \log_{10} t_{\rm E} - \log_{10} \pi_{\rm E}$ for the median parameter values from posterior samples (denoted with subscript $50$), while the Y-axis position is the probability of belonging to the `SOBH' class from the classifier using a given IFMR. Solid  straight lines represent the fast $p({\rm SOBH})$ estimation for each IFMR. The black dashed line represents the $p({\rm SOBH})$ estimation, averaging over all IFMRs (Eq. \ref{eq:p_bh_fast}). For all IFMRs, $p(\text{SOBH})$ is strongly correlated with $\log_{10} t_{\rm E} - \log_{10} \pi_{\rm E}$, though the slope of the relationship varies, with the \citet{Raithel2018} IFMR in particular returning significantly lower $p(\text{SOBH})$ values.
    }
    \label{fig:BH_linear_estimate}
\end{figure}

Fig. \ref{fig:BH_linear_estimate} shows that for the events with $p(\text{SOBH}|\boldsymbol{d}, \mathcal{G})>0.2$, there is a high correlation between $\log_{10} t_{\rm E} - \log_{10} \pi_{\rm E}$, calculated with median posterior values, and $p(\text{SOBH}|\boldsymbol{d}, \mathcal{G})$. This allows a fast approximation of the probability that an event is caused by a black hole and could be useful in real-time data streams with a large volume of events such as the Vera C. Rubin Observatory Legacy Survey of Space and Time (LSST).

We fit a straight line to all events with $\log_{10} t_{\rm E} - \log_{10} \pi_{\rm E} > 3$ for the median parameter values from posterior samples and average over all IFMR choices to obtain the relation,
\begin{equation}
    p({\rm SOBH} | x) \begin{dcases*}
     < 0.08 & for $x \leq 3$\\
    \approx 0.930 x - 2.713 & for $3 < x \leq 3.99$ \\
     \approx 1 & for $x > 3.99$,
    \end{dcases*}
\label{eq:p_bh_fast}
\end{equation}
Here, $x = \log_{10} t_{\rm E} - \log_{10} \pi_{\rm E}$. This high correlation can be understood by examining the form of $x$ and its dependence on $M_{L}$ (via Eqs. \ref{eq:tE} and \ref{eq:piE}),
\begin{equation}
\log_{10} \left(\frac{t_{\rm E}}{\pi_{\rm E}} \right) = \log_{10} \left(\frac{\kappa M_{\rm L}}{\mu_{\rm rel}} \right).
\label{eq:x}
\end{equation}
Here, $\kappa = 8.144 \text{mas} / M_{\odot}$ and $\mu_{\text{rel}}$ is the relative lens-source proper motion. Eq. (\ref{eq:x}) shows that this variable is no longer dependent on $D_{S}$ and within the logarithm is $\propto M_{L}$ compared with $t_{\rm E}\propto\sqrt{M_{L}}$. This stronger dependence on $M_{L}$ allows better discrimination for high-mass lenses such as SOBH. The effectiveness of $x$ can also be understood in terms of directions in $t_{\rm E}-\pi_{\rm E}$-space shown in Fig. \ref{fig:kdes}. $x$ is in the direction of diagonally downwards and to the right in $t_{\rm E}-\pi_{\rm E}$-space, which is the direction in which lens classes are separated. 

\subsection{Comparison with {\tt DarkLensCode}}

For the set of high-probability candidates found in Section \ref{sec:candidates}, we compared the lens classification method in this paper to the {\tt DarkLensCode}\footnote{\url{https://github.com/BHTOM-Team/DarkLensCode}} \citep{Howil2024}. While {\tt DarkLensCode} is not able to classify an event as being caused by a black hole, it can estimate lens masses, distances and the probability of the lens being dark. This method both has overlapping (i.e., black hole lenses should be dark) and complimentary (i.e., lens mass and distances) information to our classification method. 

We use the default {\tt DarkLensCode} mass function, assign weight = 0 to samples with $M_{lens} > 1000 M_\odot$, and sample for $10^6$ iterations. Where possible, we use \gaia~source proper motion and distance information. We take care to avoid contamination from the blend or astrometric lensing. In particular, where spectrophotometric distances (\texttt{distance\_gspphot} in the \gaia~source table) are available and $b_{\rm sff}$ is consistent with 1 (using the same criterion as in the \gaia~crossmatch in Section \ref{sec:candidates}), we use them along with their upper/lower bounds; otherwise we sample from the {\tt DarkLensCode} Galactic model within the range between 0 and 12 kpc for source distances. Where proper motions are available, $b_{\rm sff}$ is consistent with 1 (again to avoid issues related to blending) and the maximum approach occurred at least $2t_{\rm E}$ before the start of \gaia~observations (to avoid influence of astrometric signal), we use the \gaia~values; otherwise we sample from the Galactic model. Where available, we use the OGLE-III I-band extinction calculator\footnote{\url{https://ogle.astrouw.edu.pl/cgi-ogle/getext.py}} \citep{Nataf2013}. For events without nearby extinction gridpoints (5 out of 23 SOBH candidates) we use I-band values from \citet{Schlafly2011} extinction maps as implemented in the NED Extinction Calculator\footnote{ \url{https://ned.ipac.caltech.edu/extinction_calculator} \newline The NASA/IPAC Extragalactic Database (NED) is funded by the National Aeronautics and Space Administration and operated by the California Institute of Technology.}. We find {\tt DarkLensCode} runtimes to be highly variable depending on the auxiliary information provided, varying from $\sim$10 seconds to $\sim$10 minutes per event.

\begin{figure*}
    \centering
    \includegraphics[width=\textwidth, trim=0cm 0cm 0cm 0cm]{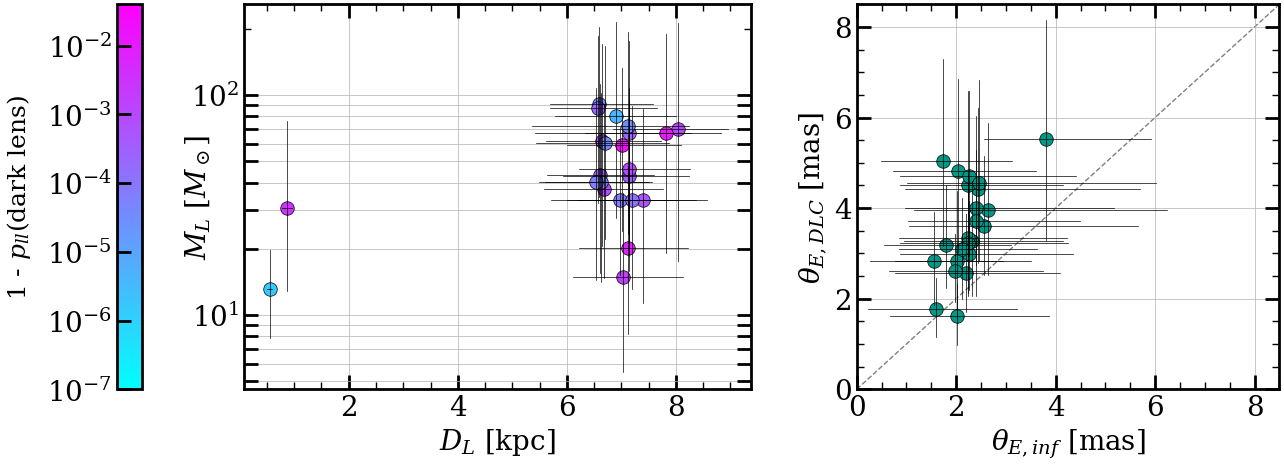}
    \caption{Results of {\tt DarkLensCode} runs for the 23 strong SOBH candidates. \textit{Left:} Lens masses and distances returned by {\tt DarkLensCode}, colored by 1 - dark lens probability (lower limit). According to {\tt DarkLensCode} results, all candidates have $>$97\% dark lens probability and high (13-91 $M_{\odot}$) masses; most candidates reside on the near side of the Galactic Bulge, with the exception of two nearby lenses at distances within 0.5-1 kpc. \textit{Right:} $\theta_{\rm E}$ inferred from the \texttt{PopSyCLE} simulation results and nested sampling (see Sec.~\ref{sec:thetaE_inference}) vs. $\theta_{\rm E}$ returned as {\tt DarkLensCode} output. Errorbars indicate 16-84 percentiles of samples. The $\theta_{\rm E, inf}$ values cluster around 2 mas, where the prior distribution from the \texttt{PopSyCLE} simulation peaks, whereas the $\theta_{\rm E, DLC}$ values are subject to less model constraints and on average higher.}
    \label{fig:dlc_comparison}
\end{figure*}

We find dark lens probabilities between 97\% and 100\% for all events, regardless of the completeness of \gaia~source information. We also find relatively high masses, with median values between 13 and 91 $M_{\odot}$; all significantly more massive than the only isolated SOBH found so far \citep{Lam2022}, and overall closer to the most massive SOBH discovered so far in our Galaxy, GaiaBH3 \citep{GaiaBH3}. Most lenses are situated on the near side of the Bulge, between 6 and 8 kpc away. The two outlying ($<$2 kpc) candidates had available \gaia~spectrophotometric distances that imposed strong constraints on the source being nearby. We present the inferred masses, distances and dark lens probabilities in Fig.~\ref{fig:dlc_comparison}.

We also compare the inferred Einstein radii $\theta_{\rm E}$ and find that \texttt{DarkLensCode} predicts higher (on average by 62\%) values; we present this comparison in Fig.~\ref{fig:dlc_comparison}.
This discrepancy can be explained as a consequence of the method. Both our inference and \texttt{DarkLensCode} use the limited information on $\theta_{\rm E}$ provided by photometric data, and fill in the missing information using their respective Galactic model assumptions. In our inference, the missing information is provided directly in the $\theta_{\rm E}$ prior constructed from a simulated population of lenses. Contrarily, {\tt DarkLensCode} uses dynamical information to infer $\theta_{\rm E}$, deriving it as $\theta_{\rm E} := t_{\rm E} \mu_{\rm rel}$. The $\mu_{\rm rel}$ values are sampled from a $\mathcal{U}(0,30)$ [mas/yr] prior and evaluated using a simple model of disk and bulge velocity distributions, thus allowing loosely constrained, high $\theta_{\rm E}$ values (i.e., there are no specific priors or constraints on the lens mass). Both methods have their strengths, as our inference is incorporating the full extent of information from the Galactic model, whereas {\tt DarkLensCode} will perform better on unpredicted lens types, e.g. intermediate-mass black holes.

\section{Classification of OB110462}

\begin{figure}[t!]
    \centering
    \includegraphics[width=\columnwidth]{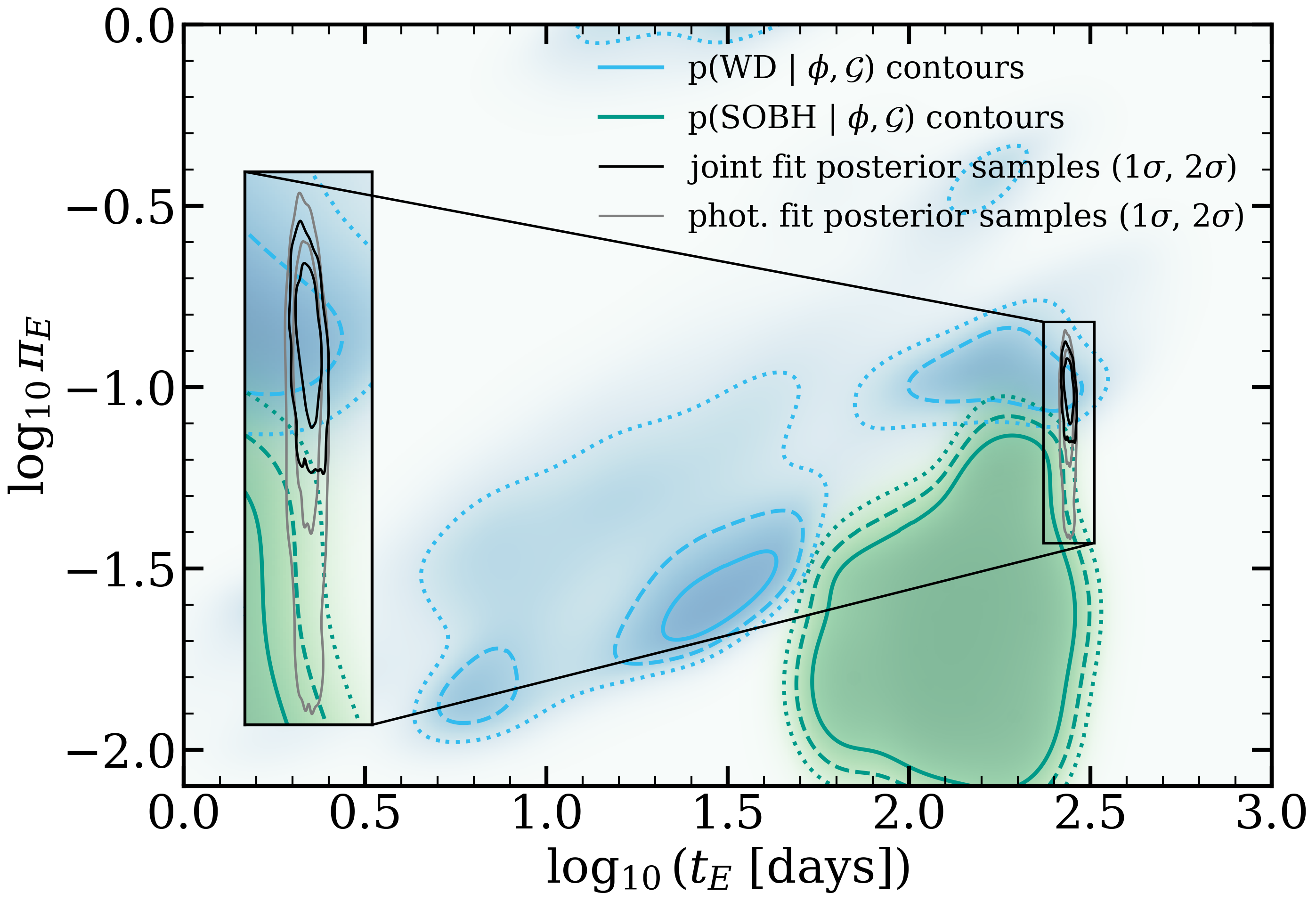}
    \caption{Classification of the only currently known isolated SOBH, OB110462 \citep{Sahu2022, Lam2022}. Background colours correspond to subplots of Fig.~\ref{fig:kdespace} for the WD and SOBH classes (using the \citet{Sukhbold2016} IFMR), overlaid with 50\% opacity each. Contours mark $p(\text{class}_L | \phi, \mathcal{G}) = {0.3, \ 0.5, \ 0.7} $ for dotted, dashed and solid lines, respectively; color coding of $\text{class}_L$ as in Fig.~\ref{fig:kdes} and thereafter. Solid contours enclose 68\% and 95\% of posterior samples from modelling of OB110462 from \citet{Lam2023}. Grey contours represent the photometry-only fit, which was used for the classification; black contours represent the joint fit using photometry and astrometry together, which further constrained the parameters. Inset shows the posterior sample distribution zoomed by a factor of 2.5. OB110462 is situated in the region of parameter space close to simulated events from both the WD and the SOBH class and showing steep gradients of $p(\text{class}_L | \phi, \mathcal{G})$.}
    \label{fig:ob110462_relative}
\end{figure}

In addition to the set of OGLE-IV microlensing events in the low cadence fields, we apply our classification method on the only known microlensing event caused by an isolated black hole -- OGLE-2011-BLG-0462, hereafter OB110462 \citep{Sahu2022, Lam2022, Lam2023}. As this work is focused on classifying microlensing events and selecting follow-up candidates based on photometry, we only use photometric data for this classification. We find that OB110462 has posterior probability of being a SOBH based on its photometric microlensing signal ranging between $p(\text{SOBH})= 0.049$ and $0.224$, depending on the Galactic model IFMR choice. The reason for this is that the $t_{\rm E}-\pi_{\rm E}$ posterior distribution for OB110462 is more consistent with the WD population (see Fig. \ref{fig:ob110462_relative}).

In the absence of astrometric microlensing data, one may use measurements of $t_{\rm E}$ and $\pi_{\rm E}$, along with a Galactic model-based assumption for proper motion, to break the mass-distance degeneracy of a photometric microlensing event. 
Thus, several SOBH candidates have been identified based on high $t_{\rm E}$ and high $\pi_{\rm E}$ \citep[e.g.,][]{Bennett2002,Poindexter:2005, Wyrzykowski2016}. 
The first astrometric microlensing follow-up programs primarily selected events from their long timescales alone, including e.g. HST Program 12322 which began monitoring OB110462 and \citet{Lu2016}.
\citet{Lam2020} showed via \texttt{PopSyCLE} that events with high $t_{\rm E}$ and low $\pi_{\rm E}$ are the best SOBH candidates in the direction of the Galactic bulge.
Prior studies targeting or including high $t_{\rm E}$, high $\pi_{\rm E}$ candidates focused on a region of parameter space not well-populated by \texttt{PopSyCLE} simulations, and thus were likely biased toward outliers that may come from any of the populations.

There are several possible interpretations for the low SOBH class probability of OB110462. Firstly, we could have just gotten lucky with this particular candidate. Although the posterior class probability of being SOBH is low, it is still on average $\approx 14$ times the prior probability of $\approx0.01$ according to the Galactic models (4.9, 27.0 and 9.2 times for Spera15, Raithel18 and Sukhbold N20 IFMRs, respectively). This means that lightcurve data does boost the chances of this event being caused by a SOBH lens. Moreover, this event was one of a set of $5$ events in a similar region of $t_{\rm E}-\pi_{\rm E}$-space that were astrometrically followed up; the other 4 were ultimately found not to be SOBHs \citep[e.g.,][]{Lam2022sup}.

Secondly, the low SOBH probability of OB110462 may suggest that the underlying Galactic models on which the classifier is based are not quite correct. Each of these Galactic models comes with its own set of assumptions (SOBH abundance, Galactic structure etc...), which, if changed, could shift the populations of SOBHs and WDs in $t_{\rm E}-\pi_{\rm E}$-space and consequently change the SOBH class probability of this event. 

If the underlying Galactic models in our classifier are correct, then OB110462 is in the tails of the SOBH distribution and outlying in $t_{\rm E}-\pi_{\rm E}$-space. Regardless of the specific interpretation of OB110462's low SOBH class probability, it is an observational outlier -- either it is an atypical SOBH and the community got lucky finding it, or it is within the astrophysical SOBH distribution, but outlying from expectations of current Galactic models.

\section{Discussion and Conclusion} \label{sec:conclusions}

We have developed and tested a new method of classifying microlensing events. Our method is effective in searching for black hole candidates, as black holes are well separated in $\log_{10} t_{\rm E}$ -- $\log_{10} \pi_{\rm E}$ space. We make a microlensing event population classifier using our methodology publicly available in the {\tt popclass}\footnote{\url{https://github.com/LLNL/popclass}} software package.\\

The method is flexible, as the underlying Galactic model can be freely modified and new lens populations (e.g. primordial black holes, free-floating planets) can be added. It also requires no additional information beyond the $t_{\rm E}$ and $\pi_{\rm E}$ posteriors (we note it is not necessary to have \textit{detectable} parallax signal, and an upper constraint is sufficient for classification). This makes the method easy to apply to large datasets, as demonstrated in this study with the classification of $\sim$10,000 OGLE-III and OGLE-IV events. In combination with being fast ($\sim$1 s for 10,000 posterior samples), lightweight and conceptually simple, this makes the method ideal for integrating into real-time processing and alerting pipelines of variability surveys.

Another advantage is not relying on the source flux fraction $b_{\rm sff}$. The most prospective fields for microlensing searches are crowded; recognising that a significant fraction of sources in those fields may have added light from close neighbours regardless of lensing, our method does not discard events with low $b_{\rm sff}$. Finally, as demonstrated in Sec.~\ref{sec:thetaE_inference}, the method can also be used to estimate $\theta_{\rm E}$ and the expected astrometric signal, which is especially useful in making decisions for allocating follow-up at $\sim 1$ mas precision.\\

We also note the limitations of our method. Firstly, it is more difficult to identify other classes of dark remnants in $\log_{10} t_{\rm E}$ -- $\log_{10} \pi_{\rm E}$ space alone; in particular, it is virtually impossible to identify high-probability neutron star candidates (see Fig.~\ref{fig:kdespace}). If those are the objects of interest, other event parameters should be used. For example, extending the parameter space to three dimensions (by adding astrometric information -- $\theta_{\rm E}$) might be helpful, which will be possible on a large scale with the \textit{Roman Space Telescope}.

As our method heavily relies on simulated events, it is only as good as the underlying Galactic model. This can be mitigated to some level with the None class, which identifies events in regions of low simulation support and may point towards missing physics. Still, it is not a failsafe indicator of incompleteness in case of overlapping populations. To minimize misclassifications, close connection between Galaxy simulations and microlensing surveys should be maintained, and the simulations -- continuously updated. In this work, we use the current state-of-the-art software for simulations including dark remnants, \texttt{PopSyCLE}.

Finally, some events may have extra information that is helpful in classifying them, e.g. spectroscopic or kinematic information about the source. This method does not include such information, which is a tradeoff to maximise simplicity and universality.\\

For the reasons outlined above, we conclude this method is complimentary to existing methods. Our classifier is well-positioned to work as an initial filter, picking out the best black hole candidates from a large dataset and optimizing the allocation of follow-up resources. Those candidates can then be treated on an individual basis, including adding auxiliary information.

We applied this method to search for black hole candidates in OGLE-III and OGLE-IV data. We find 23 events that we classify as strong black hole candidates, i.e. more likely to be a black hole than any other class regardless of the IMFR used. As all those events are archival and happened between 2002 and 2017, the possibility for analysis is limited. Some of our candidates have \gaia~data collected during amplification and can be revisited in the time-series dataset of the upcoming \gaia~Data Release 4. Analysis with the {\tt DarkLensCode} software ascribes very high ($>97\%$) probabilities of being a dark remnant and high (13-91 $M_\odot$) masses to all of our strong black hole candidates.  
We also predict $\theta_{\rm E}$ from the posterior predictive distribution and the expected astrometric signal. Based on this method, we could have made follow-up decisions to observe candidates most likely to be detected astrometrically. We note that a fast reaction to follow up at the moment of highest astrometric deviation ($u = \sqrt{2}$) may be prioritised over having a maximally precise event posterior (return to baseline), as astrometric signal from the black hole candidates even at maximum is expected to be on the verge of current detection possibilities.

We find that the black hole classifications in particular are heavily impacted by the underlying IFMR, with the Raithel18 simulation run yielding significantly less black hole candidates than the other two. This also strongly limits the size of our black hole candidate sample. Future surveys such as \textit{Roman Space Telescope} will populate the $\log_{10} t_{\rm E} - \log_{10} \pi_{\rm E}$ space with tightly constrained datapoints. By working backwards from lensing event parameter distributions, IFMRs can be constrained, and the underlying models of stellar evolution and supernova physics verified. This will be effective especially if the astrophysical class label is available independently, e.g. from astrometric mass measurement.

Looking forward, we anticipate an interesting and intense time for microlensing. Some ongoing astrometric microlensing SOBH searches have recently begun to focus on high $t_E$, low $\pi_E$ candidates (e.g. JWST DD Program 6777), and should have higher chances of success, in addition to providing valuable insights on this selection method’s effectiveness and the accuracy of the underlying Galactic model. The increasing yield of events will make it critical to classify them in the most efficient way with minimal human intervention. The {\it Legacy Survey of Space and Time} at the {\it Vera C. Rubin Observatory}, which will detect thousands of photometric microlensing events over a wide field with high-cadence observations, presents an ideal case for this classifier which could be integrated into Target and Observation Managers \citep[e.g.,][]{Street2018, vanderWalt2019, Coulter2022, Coulter2023}. The classifier can also be applied to \textit{Roman Space Telescope} data, with a possible extension to three dimensions with astrometry. Applying our method to the upcoming large microlensing event datasets may yield a substantial sample of isolated SOBHs and answer long-standing questions about this elusive population.

\section*{Acknowledgements}

We thank George Chapline and Simeon Bird for many useful discussions on this work. We also thank Casey Lam for the photometric posterior distributions of OB110462. This work was performed under the auspices of the U.S. Department of Energy by Lawrence Livermore National Laboratory under Contract DE-AC52-07NA27344. The document number is LLNL-JRNL-870480. This work was supported by the LLNL-LDRD Program under Project 22-ERD-037. This document was prepared as an account of work sponsored by an agency of the United States government. 
Neither the United States government nor Lawrence Livermore National Security, LLC, nor any of their employees makes any warranty, expressed or implied, or assumes any legal liability or responsibility for the accuracy, completeness, or usefulness of any information, apparatus, product, or process disclosed, or represents that its use would not infringe privately owned rights. 
Reference herein to any specific commercial product, process, or service by trade name, trademark, manufacturer, or otherwise does not necessarily constitute or imply its endorsement, recommendation, or favoring by the United States government or Lawrence Livermore National Security, LLC. ZK acknowledges support from the 2024 LLNL Data Science Summer Institute and is a Fellow of the International Max Planck Research School for Astronomy and Cosmic Physics at the University of Heidelberg (IMPRS-HD). N.S.A., M.J.H., and J.R.L. acknowledge support from the National Science Foundation under grant No. 1909641 and the Heising-Simons Foundation under grant No. 2022-3542.
M.F.H. was supported by a NASA FINESST grant No. ASTRO20-0022.

\vspace{5mm}

\software{This research has made use of NASA's Astrophysics Data System Bibliographic Services. \texttt{NumPy} \citep{Harris2020}, \texttt{SciPy} \citep{Virtanen2020}, \texttt{Matplotlib} \citep{Hunter2007}, \texttt{Astropy} \citep{astropy:2013,astropy:2018,astropy:2022}, \texttt{PopSyCLE} \citep{Lam2020}, \texttt{Galaxia} \citep{Sharma2011}, \texttt{SPISEA} \citep{Hosek2020}, \texttt{scikit-learn} \citep{Pedregosa2011}, {\tt dynesty} \citep{DYNESTY}}

\appendix

\section{Comparison of classification with and without the None class}\label{app:compare}

\begin{figure}
    \centering
    \includegraphics[width=\textwidth, trim=0cm 0cm 0cm 0cm]{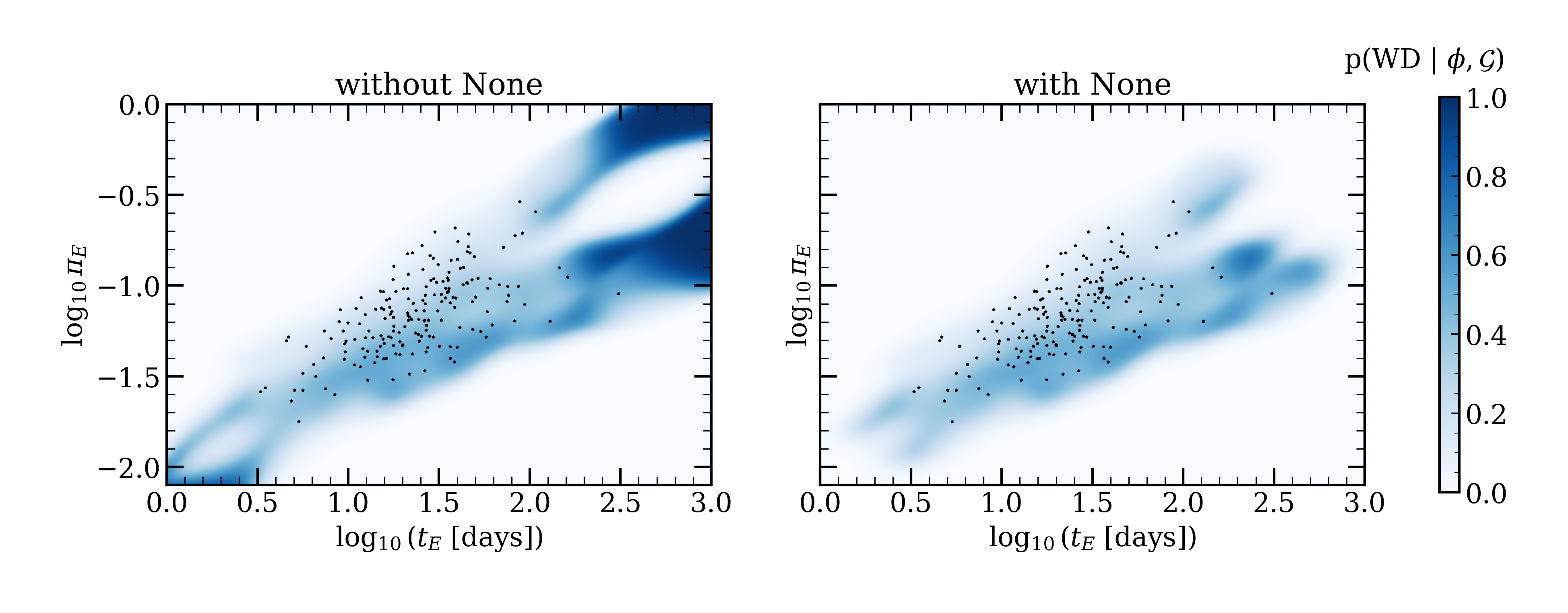}
    \caption{Relative classification probabilities for the WD class for points on 1000x1000 grid, using the Spera15 IFMR, without (left) and with (right) the None class added to the classifier. Black points represent simulated events belonging to the class. The None class effectively erases the high $p(\text{WD}|\phi, \mathcal{G})$ patterns in regions of low simulation support, leaving only those in proximity to simulated white dwarf lens events.}
    \label{fig:kdespace_old}
\end{figure}

In the additional Figure \ref{fig:kdespace_old}, we demonstrate an example of artificial regions of very high relative probability in the parts of parameter space not covered by simulated events. Without accounting for this effect, paradoxically, e.g. a lens classified as $p(\text{WD} | \mathbf{d}, \mathcal{G}) = 99\%$ should be treated as a less reliable white dwarf candidate than a $p(\text{WD} | \mathbf{d}, \mathcal{G}) = 50\%$ one, as there are no white dwarfs in the simulation output that would match its parameters. The simulation used in this figure uses the Spera15 IFMR, which exhibits especially striking $\approx 100\%$ relative probability features in regions of no information before adding the None class. However, all simulation runs are affected.

\section{Simulation fields}\label{app:sim}

\begin{table}[]
    \centering
    \hspace*{-2cm}\begin{tabular}{p {4cm} p{2.5cm} p{2.2cm} p{1.7cm} p{1.7cm} p{2.5cm}}
        \hline \hline
        Field ($l$, $b$) & $N$(star) & $N$(WD) & $N$(NS) & $N$(SOBH) & $N_{\rm total}$ \\ \hline(-5.6069$^{\circ}$, -2.0233$^{\circ}$) & (46, 39, 36) & (7, 5, 6) & (0, 0, 1) & (2, 0, 0) & (55, 44, 43) \\
(-4.2794$^{\circ}$, -5.4419$^{\circ}$) & (31, 30, 20) & (4, 1, 3) & (1, 0, 3) & (0, 0, 1) & (36, 31, 27) \\
(-4.2223$^{\circ}$, -6.8055$^{\circ}$) & (5, 13, 3) & (1, 2, 2) & (1, 0, 0) & (0, 0, 0) & (7, 15, 5) \\
(-4.2100$^{\circ}$, 4.9609$^{\circ}$) & (26, 21, 27) & (3, 0, 3) & (0, 2, 1) & (0, 0, 1) & (29, 23, 32) \\
(-3.2832$^{\circ}$, -3.4735$^{\circ}$) & (91, 82, 76) & (11, 10, 17) & (0, 2, 1) & (3, 2, 1) & (105, 96, 95) \\
(-3.2058$^{\circ}$, -4.8329$^{\circ}$) & (58, 50, 58) & (8, 8, 4) & (1, 0, 1) & (0, 0, 0) & (67, 58, 63) \\
(-1.9286$^{\circ}$, 1.3682$^{\circ}$) & (42, 64, 50) & (3, 10, 6) & (0, 1, 0) & (0, 1, 0) & (45, 76, 56) \\
(-1.0641$^{\circ}$, -3.6101$^{\circ}$) & (130, 116, 114) & (22, 22, 20) & (0, 2, 3) & (0, 2, 1) & (152, 142, 138) \\
(-0.9534$^{\circ}$, -6.3377$^{\circ}$) & (34, 21, 26) & (4, 2, 3) & (0, 0, 1) & (0, 1, 0) & (38, 24, 30) \\
(0.7819$^{\circ}$, 1.6875$^{\circ}$) & (198, 190, 193) & (36, 29, 21) & (3, 3, 4) & (10, 3, 3) & (247, 225, 221) \\
(1.1399$^{\circ}$, -3.7432$^{\circ}$) & (229, 179, 150) & (23, 25, 24) & (2, 2, 1) & (3, 2, 4) & (257, 208, 179) \\
(3.3316$^{\circ}$, -3.8823$^{\circ}$) & (120, 118, 120) & (16, 13, 12) & (1, 4, 0) & (1, 2, 1) & (138, 137, 133) \\
(3.5176$^{\circ}$, 3.5577$^{\circ}$) & (66, 60, 61) & (9, 3, 6) & (0, 1, 0) & (1, 3, 1) & (76, 67, 68) \\
(3.6341$^{\circ}$, 2.1945$^{\circ}$) & (45, 57, 57) & (6, 8, 7) & (3, 1, 1) & (2, 1, 0) & (56, 67, 65) \\
(4.4046$^{\circ}$, -3.2761$^{\circ}$) & (84, 79, 68) & (8, 8, 10) & (0, 0, 1) & (1, 1, 2) & (93, 88, 81) \\
(4.6747$^{\circ}$, 2.8534$^{\circ}$) & (84, 86, 74) & (8, 6, 16) & (0, 2, 1) & (1, 1, 2) & (93, 95, 93) \\
(5.4762$^{\circ}$, -2.6684$^{\circ}$) & (46, 38, 38) & (3, 10, 4) & (1, 1, 1) & (1, 0, 0) & (51, 49, 43) \\
(5.6025$^{\circ}$, 4.8747$^{\circ}$) & (19, 14, 17) & (0, 2, 3) & (1, 1, 0) & (0, 0, 0) & (20, 17, 20) \\
(5.8168$^{\circ}$, 2.1491$^{\circ}$) & (49, 53, 46) & (9, 4, 8) & (0, 1, 1) & (1, 1, 0) & (59, 59, 55) \\
(6.6383$^{\circ}$, -4.8152$^{\circ}$) & (29, 24, 21) & (4, 2, 3) & (0, 1, 0) & (0, 0, 0) & (33, 27, 24) \\

        \hline
    \end{tabular}
    \caption{Event counts in the 20 simulated fields. The first column shows field centerpoints (all fields are circles with an area of 0.3 deg$^2$). The following columns contain numbers of events in the corresponding fields belonging to a given lens class in the Spera15, Raithel18 and Sukhbold N20 simulations, respectively.
}
    \label{tab:sim_stats}
\end{table}

\bibliography{ref}{}
\bibliographystyle{aasjournal}

\end{document}